\documentclass[reprint,aps,prapplied,longbibliography,twocolumn,10pt,superscriptaddress,nobibnotes]{revtex4-2}
\usepackage{graphicx}	
\usepackage{subfigure}
\usepackage{dcolumn}  					
\usepackage{bm} 
\usepackage{siunitx}

\begin{document}

\title{Long-distance chronometric leveling with a portable optical clock}

\author{J.~Grotti}
\author{I.~Nosske}
\author{S.~B.~Koller}
\author{S.~Herbers}
\affiliation{Physikalisch-Technische Bundesanstalt, Bundesallee 100, 38116 
	Braunschweig, Germany}

\author{H.~Denker}
\author{L.~Timmen}
\affiliation{Institut f\"ur Erdmessung, Leibniz Universit\"at Hannover (LUH), Schneiderberg 50, 30167 Hannover, Germany}

\author{G.~Vishnyakova}
\affiliation{Max-Planck-Institut f\"ur Quantenoptik, Hans-Kopfermann-Stra{\ss}e 1, 85748 Garching, Germany}

\author{G.~Grosche}
\author{T.~Waterholter}
\author{A.~Kuhl}
\author{S.~Koke}
\author{E.~Benkler}
\affiliation{Physikalisch-Technische Bundesanstalt, Bundesallee 100, 38116 Braunschweig, Germany}

\author{M.~Giunta}
\affiliation{Max-Planck-Institut f\"ur Quantenoptik, Hans-Kopfermann-Stra{\ss}e 1, 85748 Garching, Germany}
\affiliation{Menlo Systems GmbH, Bunsenstra{\ss}e 5, 82152 Martinsried, Germany}

\author{L.~Maisenbacher}
\author{A.~Matveev}
\affiliation{Max-Planck-Institut f\"ur Quantenoptik, Hans-Kopfermann-Stra{\ss}e 1, 85748 Garching, Germany}

\author{S.~D\"orscher}
\author{R.~Schwarz}
\author{A.~Al-Masoudi}
\affiliation{Physikalisch-Technische Bundesanstalt, Bundesallee 100, 38116 Braunschweig, Germany}

\author{T.~W.~H\"ansch}
\author{Th.~Udem}
\affiliation{Max-Planck-Institut f\"ur Quantenoptik, Hans-Kopfermann-Stra{\ss}e 1, 85748 Garching, Germany}

\author{R.~Holzwarth}
\affiliation{Max-Planck-Institut f\"ur Quantenoptik, Hans-Kopfermann-Stra{\ss}e 1, 85748 Garching, Germany}
\affiliation{Menlo Systems GmbH, Bunsenstra{\ss}e 5, 82152 Martinsried, Germany}

\author{C.~Lisdat}
\email{christian.lisdat@ptb.de}
\affiliation{Physikalisch-Technische Bundesanstalt, Bundesallee 100, 38116 Braunschweig, Germany}

\begin{abstract}
We have measured the geopotential difference between two locations separated by 457~km by comparison of two optical lattice clocks via an interferometric fiber link, utilizing the gravitational redshift of the clock transition frequency. 
The $^{87}$Sr clocks have been compared side-by-side before and after one of the clocks was moved to the remote location. 
The chronometrically measured geopotential difference of $3918.1(2.6)\,\mathrm{m^2 \, s^{-2}}$ agrees with an independent geodetic determination of $3915.88(0.30)\,\mathrm{m^2 \, s^{-2}}$.
The uncertainty of the chronometric geopotential difference is equivalent to an uncertainty of $27~\mathrm{cm}$ in height.
\end{abstract}
\maketitle

\noindent According to general relativity, clocks located in gravity potentials that differ by $\Delta U$ experience a tick rate difference that is given in relative units by $\Delta U/c^2$ in the first (Newtonian) order, which is known as the gravitational redshift.
While this is a small effect, modern optical atomic clocks with their outstanding fractional uncertainty of $\sim 10^{-18}$~\cite{ush15, hun16, bel21, hua22} can resolve a potential difference corresponding to a height difference of $\Delta h \sim 1$~cm close to the geoid, where $\Delta h = \Delta U/g$ for a homogeneous gravity acceleration $g$. Furthermore, the frequency of two atomic clocks can be compared via an interferometric fiber link (IFL), which supports sub-$10^{-19}$ accuracy over hundreds of kilometers \cite{pre12, lis16, can21, cli20a, aka20, sch22a}.
Hence, determination of gravitational redshift differences with clocks can be utilized in geodesy to revolutionize geopotential and
physical height determination capabilities. This method is known as ``chronometric leveling'' \cite{ver83, bje85, meh18}. Experiments in this context have been conducted on short \cite{cho10a, tak20, hua20, liu23, bot22, zhe23b} and long \cite{lis16, tak16, gro18a} distances. 

Physical heights relate directly to the geopotential (a physical quantity), in contrast to the purely geometrically defined ellipsoidal heights (distance of a point from a given reference ellipsoid) from Global Navigation Satellite Systems (GNSS) measurements \cite{san21a}. The most widely known geodetic technique to derive physical heights is geometric leveling (also known as spirit leveling) in combination with local gravity measurements.
However, this method becomes fairly labor-intensive for distances of hundreds of kilometers and is prone to the accumulation of errors, which may reach several decimeters over a 1000-km distance. 
Alternatively, absolute geopotential values with uncertainties at the centimeter level can be obtained using the ``GNSS/geoid approach'' \cite{den17}. 
It combines ellipsoidal heights from GNSS with an accurate gravity field model (geoid) with high spatial resolution of a few kilometers.
The latter is obtained by refining satellite gravity field models with terrestrial gravity measurements and terrain data.
Compared to these two established geodetic methods for measuring geopotential differences, the main advantage of chronometric leveling lies in its ability to directly measure potential differences with mostly distance-independent uncertainty.

Previously, extremely small redshifts have been resolved with optical clocks \cite{bot22, zhe23b} by differential observations in a single apparatus. Long-distance geodetic measurements require independent clocks, however. 
While stationary optical clocks have been compared over long distances \cite{lis16, tak16}, chronometric leveling requires an additional side-by-side comparison to establish the frequency ratio of the clock apparatuses themselves, i.e., to measure any potential frequency shifts other than the gravitational redshift. Hence, at least one of the clocks needs to be transported.
Chronometric leveling with a transportable optical clock has first been realized with a $^{87}$Sr lattice clock \cite{kol17} between Modane (France) and Torino (Italy), albeit with a moderate fractional frequency uncertainty of $1.9 \times 10^{-15}$ \cite{gro18a}.
Today, several groups operate transportable optical clocks \cite{ori18, gel20, hua20, ohm21, fas21a, stu21, guo21, liu23, zen23}. 
In a local area height measurement, an accuracy of 4~cm has been demonstrated \cite{tak20}.

In this work, we demonstrate long-distance chronometric leveling with two optical clocks over a distance of several 100 km with a frequency uncertainty at the low $10^{-17}$ level for the first time (corresponding to a height resolution of a few decimeters). The two connected sites are the Physikalisch-Technische Bundesanstalt (PTB) at Braunschweig, Germany and the Max-Planck Institute of Quantum Optics (MPQ) at Garching, Germany that are 457~km (linear distance) apart and have a height difference of 400~m. 
They are connected via a 940-km-long IFL (see Fig.~\ref{fig:FiberLink}). 
PTB's transportable $^{87}$Sr lattice clock Sr2 \cite{kol17} in its air-conditioned car trailer was compared against the stationary $^{87}$Sr lattice clock Sr1 \cite{fal14, sch20d} before, during, and after its stay at MPQ. 
The local clock comparisons at PTB ensure that the frequency ratio between colocated clocks remains unaffected by the transport and is known.
Only then a frequency shift observed at the remote site can be attributed to the gravitational redshift.
To verify our chronometric leveling result, we compare it with the result of the GNSS/geoid approach and find good agreement.

{\begin{figure}[tb]
\centering
\includegraphics[width=0.7\linewidth]{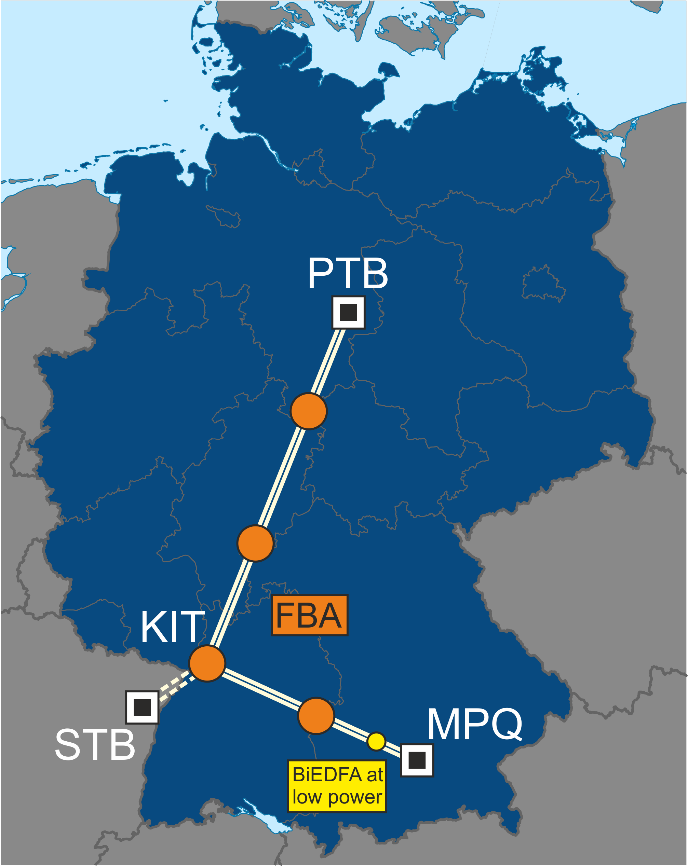}
\caption{(Color online) The 940-km-long IFL between PTB and MPQ, shown on a map of Germany (in blue). 
Due to attenuation of the link laser light, four fiber Brillouin amplifiers (FBAs) \cite{rau15} and one bidirectional erbium-doped fiber amplifier (BiEDFA) have been installed along its path. At Karlsruhe Institute of Technology (KIT), the IFL from PTB branches to MPQ and Strasbourg (STB).}
\label{fig:FiberLink}
\end{figure}}

 \begin{figure}[tb]
    \centering
    \includegraphics[width=1\linewidth]{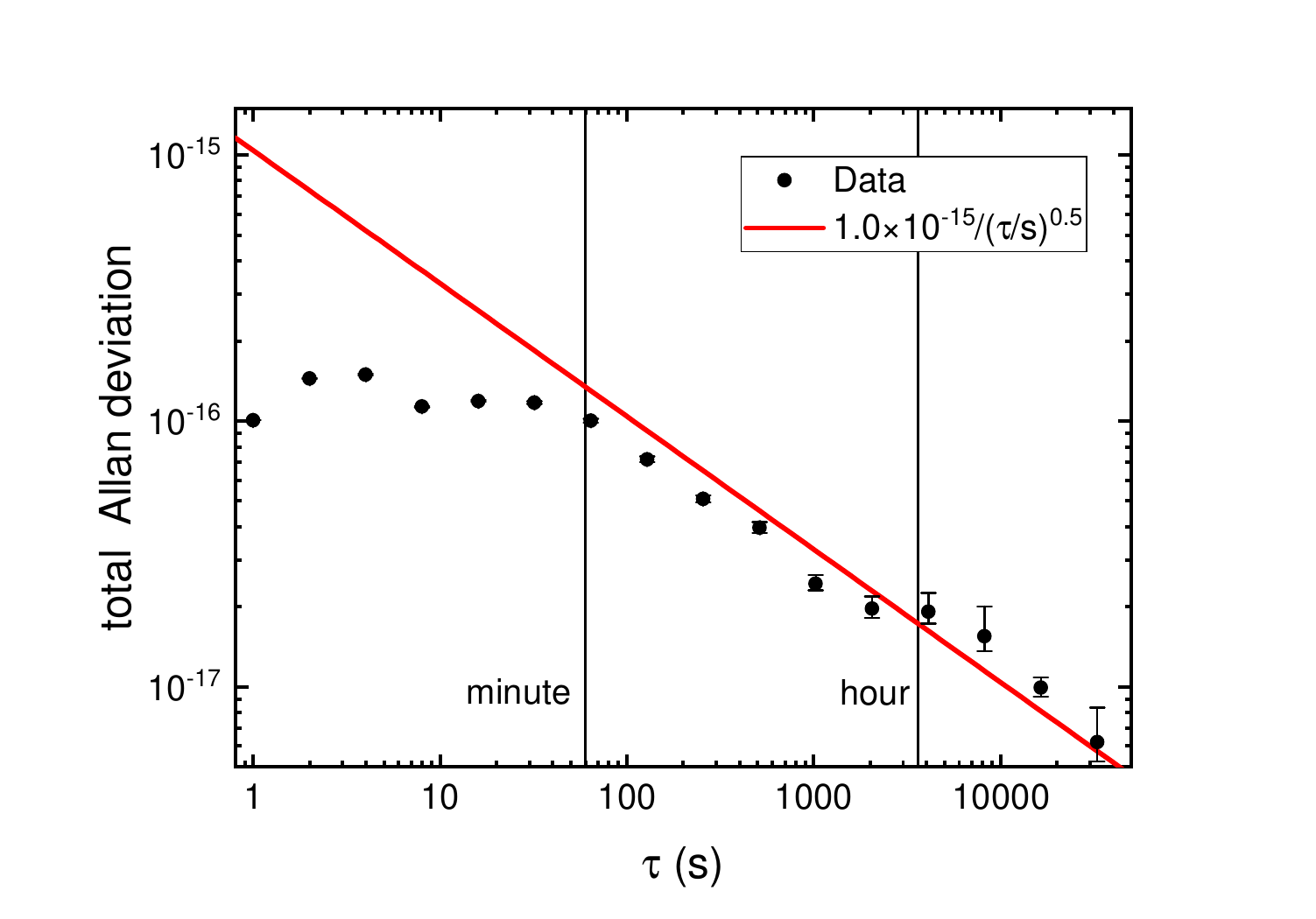}
    \caption{(Color online) Total Allan deviation of the concatenated data of the two local clock comparisons as a function of the averaging time $\tau$. The red line indicates the instability and is a $\propto \tau^{-1/2}$ fit to points with $\tau > 100~\mathrm{s}$.}
    \label{fig:ADEVloc}
\end{figure}
 
 Our data are recorded and analyzed as follows:
 All beat notes between the different lasers are counted with synchronized, dead-time-free counters in phase-averaging ($\Lambda$) mode and with an averaging time of 1~s \cite{kra04a,ben15}. 
 Valid counter data, as determined from auxiliary clock monitoring data, are used to calculate the difference of the measured clock transition frequencies $\Delta \nu^\mathrm{clock} = \nu_\mathrm{Sr2} - \nu_\mathrm{Sr1}$ in one-second intervals.
 The arithmetic mean of $\Delta \nu^\mathrm{clock}$ on each day is denoted by $\Delta \nu^\mathrm{clock}_\mathrm{day}$.

\begin{table}[t]
	\begin{tabular}{l | S | S | S | S} 
		\hline\hline
		uncertainty  & \multicolumn{1}{c|}{Sr1} & \multicolumn{3}{c}{Sr2}  \\
		contribution & \multicolumn{1}{c|}{} & \multicolumn{1}{c|}{(PTB~I)} & \multicolumn{1}{c|}{(MPQ)} &  \multicolumn{1}{c}{(PTB~II)} \\
		\hline
		Lattice light shift &  3.0 &  8.3 & 19.8 & 35.5 \\
		BBR ambient         & 12.8 &  6.3 & 19.5 & 15.6 \\	
		BBR oven            &  1.2 &  0	  &  0   & 0    \\	
		2nd order Zeeman    &  1.0 &  1.9 &  1.5 & 1.7  \\	
		Cold collisions     &  0.4 &  7.6 & 11.8 & 0.9  \\
		Background gas      &  2.3 &  3.8 &  3.8 & 3.8  \\
		Servo error         &  0.1 &  0   &  2.2 & 0    \\
		Tunneling           &  6.8 &  0   &   0  & 0    \\
		DC Stark shift      &  0.7 & <0.1 & <0.1 & <0.1 \\
		Probe light shift   & <0.1 &  0.2 &  0.2 &  0.2 \\
		Line pulling        & <0.1 & <0.1 & <0.1 & <0.1 \\
		\hline 
		Total               & 15.1 & 13.6 & 30.5 & 39.0 \\
		\hline
		\hline
	\end{tabular}
	\caption{Typical contributions to the fractional systematic clock uncertainties of Sr1 and Sr2 in $10^{-18}$. BBR stands for blackbody radiation. For typical frequency shifts related to the effects, refer to \cite{sch20d} and \cite{kol17} for Sr1 and Sr2, respectively. For Sr2, we differentiate between the two local (PTB~I, PTB~II) and the remote (MPQ) campaigns.}
	\label{tab:unc}
\end{table}

For the two local campaigns at PTB, we define the daily average clock offset $\Delta \nu_\mathrm{day}^\mathrm{loc}$ by

\begin{eqnarray}\label{eq:DailyRatio}
\Delta \nu_\mathrm{day}^\mathrm{loc} = \Delta\nu^\mathrm{clock}_\mathrm{day} + \mathrm{cor}^\mathrm{Sr2}_\mathrm{day} - \mathrm{cor}^\mathrm{Sr1}_\mathrm{day} - \Delta \nu^\mathrm{grav}_\mathrm{day}~,
\end{eqnarray}

\noindent where $\mathrm{cor}^\mathrm{Sr1}_\mathrm{day}$ and $\mathrm{cor}^\mathrm{Sr2}_\mathrm{day}$ are the frequency corrections of the respective clocks due to known atomic frequency shifts for a given day of measurement.
The gravitational redshift difference is given by $\Delta\nu^\mathrm{grav}_\mathrm{day} =
(\nu_0 \, g_\mathrm{PTB}/c^2) \, \Delta h_\mathrm{day}$, with the clock transition frequency $\nu_0 \approx 429.228 \, \mathrm{THz}$, the speed of light $c$, the local gravity acceleration $g_\mathrm{PTB} \approx 9.813 \, \mathrm{m \, s^{-2}}$, and the height difference between the two clocks $\Delta h_\mathrm{day} = h_\mathrm{Sr2,day} - h_\mathrm{Sr1}$. 
The height $ h_\mathrm{Sr2,day}$ is constant within each local campaign at PTB, but differs slightly between both campaigns, as the transportable clock trailer was installed at slightly different heights.
For the local clock comparisons before and after the transport to MPQ, the height difference $\Delta h_\mathrm{day}$ of the clocks' atomic clouds was measured by geometric leveling to be $-104(5) \, \mathrm{mm}$ and $-166(11) \, \mathrm{mm}$, respectively.

The uncertainty $u(\Delta \nu_\mathrm{day}^\mathrm{loc})$ of the daily data $\Delta \nu_\mathrm{day}^\mathrm{loc}$ is calculated by summing the uncertainties of the individual terms of Eq. (\ref{eq:DailyRatio}) in quadrature. 
The statistical uncertainty contribution to $\Delta \nu^\mathrm{clock}_\mathrm{day}$ is estimated from the Allan deviation of the daily measurement extrapolated to the measurement duration.
The Allan deviation of the concatenated data set of all local measurements is shown in Fig.~\ref{fig:ADEVloc}.
The uncertainties of $\mathrm{cor}^\mathrm{Sr1}_\mathrm{day}$ and $\mathrm{cor}^\mathrm{Sr2}_\mathrm{day}$ are the systematic uncertainties of the clocks Sr1 and Sr2.  
Combined, they contribute about $2 \times 10^{-17}$ and $4 \times 10^{-17}$ in fractional units of $\nu_0$ to the daily averages of the first and second local comparison campaign, respectively.
In comparison to \cite{kol17}, the uncertainty of Sr2 was reduced owing to an extended investigation of the cold-collision shift and a combined evaluation of the lattice light shifts of different orders, using the model from Ref. \cite{ush18}. 
In both cases, the main reduction of uncertainty was caused by a lower statistical uncertainty of the underlying data sets compared to \cite{kol17}.
The individual systematic uncertainty contributions of both clocks are listed in Tab.~\ref{tab:unc}.
Details are discussed in the Supplemental Material \cite{SupplementalMaterial} and references therein \cite{gum08, kol17, mid12a, nic15, lis21a, ush18, alm15, boy07a, doe18, mun21, bai07, shi15, bot19, alv19, fal11, lem05}.

We compute the local clock offset $\overline{\Delta \nu}_\mathrm{loc}$ as the average over all local measurement days $\Delta \nu_\mathrm{day}^\mathrm{loc}$ with weights $1 / u(\Delta \nu_\mathrm{day}^\mathrm{loc})^2$.
In order to account for the increased scatter of the daily clock comparison data during the first local campaign, we multiply its daily uncertainties by the data set's Birge ratio \cite{kac02}, which is about 1.7.
The uncertainty $u(\overline{\Delta \nu}_\mathrm{loc})$ is determined taking into account correlations of systematic clock corrections between the different daily measurements \cite{gum08}.
We consider the systematic corrections of each clock with the exception of the lattice light shift to be fully correlated between all measurement campaigns.
The lattice light shift is remeasured in each campaign and thus is only correlated within a campaign.
Similarly, the redshift correction for each local campaign is fully correlated within each campaign.
No further correlations are taken into account.
For more information refer to the Supplemental Material \cite{SupplementalMaterial}.
{\begin{figure}[tb]
\centering
\includegraphics[width=1\linewidth]{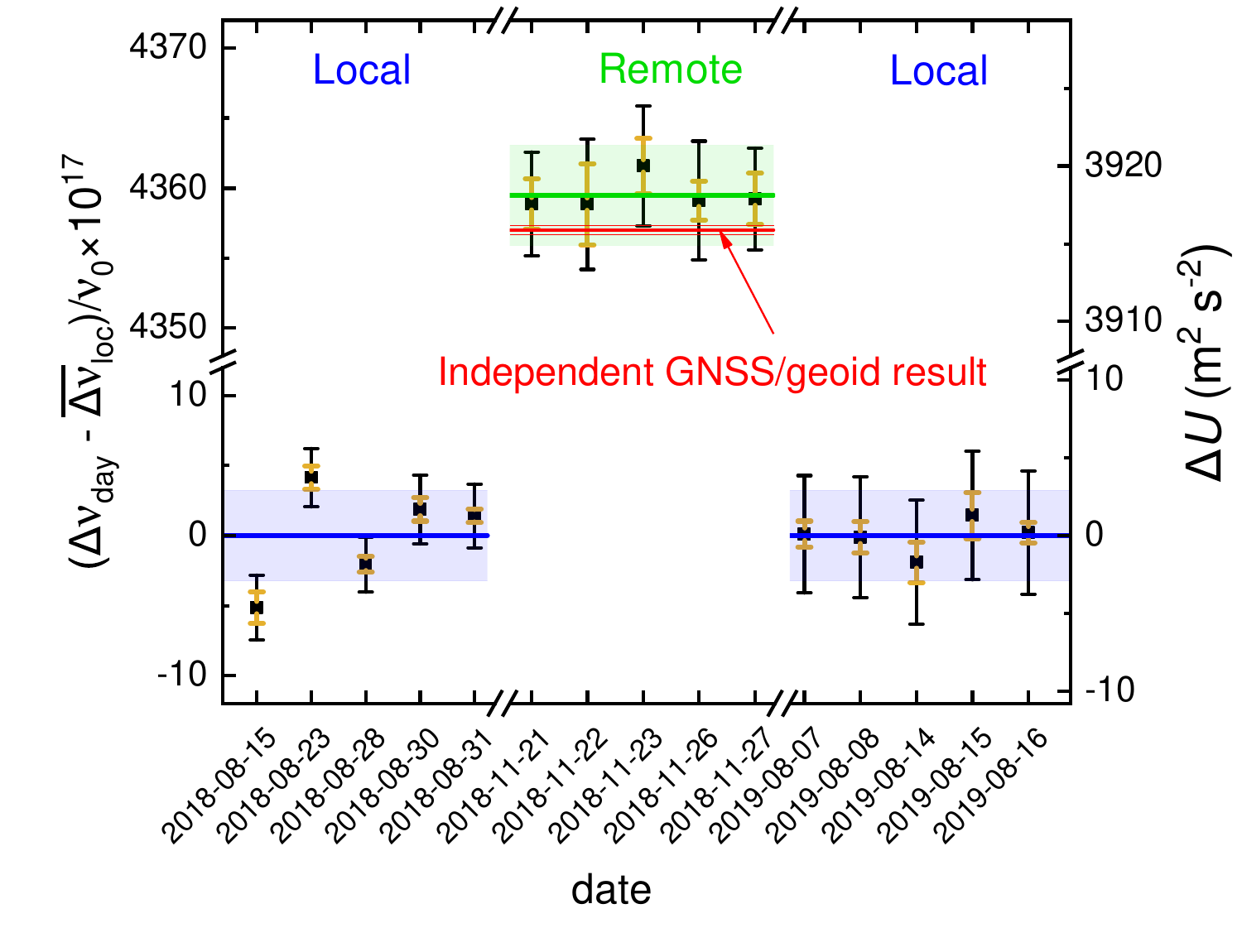}
\caption{(Color online) Daily frequency differences between the transportable clock (Sr2) and the stationary clock (Sr1) during the local and remote clock comparisons, corrected for the measured local clock offset $\overline{\Delta\nu}_\mathrm{loc}/\nu_0 = 50 \times 10^{-18}$. The plotted data include the clocks' frequency corrections, as well as corrections according to the small geopotential differences during the local measurements. The blue (green) line and shaded areas represent the mean value and uncertainty of the local (remote) clock comparison. The golden error bars represent daily statistical uncertainties, black error bars show the daily total uncertainties. The red line represents the geodetic result obtained by the combination of GNSS data and geoid modeling \cite{den17}.}
\label{fig:FreqRatios}
\end{figure}}
This results in a measured fractional local clock offset $\overline{\Delta \nu}_\mathrm{loc}/\nu_0 = 50(32) \times 10^{-18}$, 
where the combined estimated measurement uncertainty largely arises from the systematic uncertainties of the two clocks.
The fact that we observe a significant frequency offset between the clocks in the local comparisons is unexpected, because both clocks operate on the same transition $^1\mathrm{S}_0\mbox{--}{^3\mathrm{P}}_0$ of the same isotope $^{87}$Sr.
The reason for this remains unclear, even though detailed systematic investigations of the systematic frequency shifts have been performed.
However, equal offsets have been observed before and after the transfer of Sr2 to MPQ (Figs.~\ref{fig:ADEVloc} and \ref{fig:FreqRatios}).
It is therefore reasonable to assume that the offset during the remote clock comparison at MPQ had the same value as observed locally.
Thus, we regard $\overline{\Delta \nu}_\mathrm{loc}$ as a calibration for the clock's frequency ratio, as it is required if two clocks of different types with an unknown frequency ratio are used for chronometric leveling.

After the transportable clock was moved to MPQ, within one week ultracold $^{87}$Sr atoms were trapped in the optical lattice and the clock transition was found. Optical frequency combs were placed at both ends for phase-coherent frequency conversion \cite{ben19, giu19} between the clock interrogation lasers at 698~nm and the 1542-nm laser that is used for frequency dissemination over the IFL.

Daily frequency offsets during the remote comparison are calculated in the same way as the local ones in Eq.~(\ref{eq:DailyRatio}), except for the to-be-inferred gravity potential term:

\begin{equation}
\Delta \nu_\mathrm{day}^\mathrm{rem} = \Delta\nu^\mathrm{clock}_\mathrm{day} + \mathrm{cor}^\mathrm{Sr2}_\mathrm{day} - \mathrm{cor}^\mathrm{Sr1}_\mathrm{day}
\end{equation}

\noindent 
The average remote clock offset $\overline{\Delta \nu}_\mathrm{rem}$ is again calculated as the weighted average of the daily ratios.
Its uncertainty takes into account correlations of various systematic corrections as discussed above for the local comparisons.
We find $\overline{\Delta \nu}_\mathrm{rem}/\nu_0 = 43\,645(36) \times 10^{-18}$. 
The daily frequency offsets of all three clock comparisons, two local and one remote, are shown in Fig.~\ref{fig:FreqRatios}.
The total Allan deviation, shown in Fig.~\ref{fig:ADEVs}, expresses the instability of the remote clock comparison.
The IFL contribution only dominates the instability of the clock comparison data up to averaging times $\tau$ of about $\tau = 2~\mathrm{s}$. 
To the best of our knowledge, this measurement constitutes the most stable transportable clock operation published so far.
The instability is slightly higher in the remote clock comparison ($1.7 \times 10^{-15}/\sqrt{\tau / \mathrm{s}}$, Fig.~\ref{fig:ADEVs}) than in the local one ($1.0 \times 10^{-15}/\sqrt{\tau / \mathrm{s}}$, Fig.~\ref{fig:ADEVloc}) as the transportable clock laser \cite{hae20} was pre-stabilized to the ultrastable lasers at PTB \cite{hae15a, mat17a} during the local comparisons.

Taking the local clock offset, $\overline{\Delta \nu}_\mathrm{loc}$, into account, the chronometrically measured geopotential difference $\Delta U_\mathrm{chron}$ between Sr2 at MPQ and Sr1 at PTB is then given by
\begin{equation}
\Delta U_\mathrm{chron} = \frac{c^2}{\nu_0} \left( \overline{\Delta \nu}_\mathrm{rem} - \overline{\Delta \nu}_\mathrm{loc} \right) \mbox{.}
\end{equation}
Its measured value is $3918.1(2.6) \, \mathrm{m^2 \, s^{-2}}$, which agrees with the independent, geodetic determination of $3915.88(0.30) \, \mathrm{m^2 \, s^{-2}}$ (Fig.~\ref{fig:FreqRatios}).
The uncertainty of $\Delta U_\mathrm{chron}$ is determined by applying the methods described above to incorporate correlations.

The geodetic potential difference between reference points at PTB \cite{rie20} and at MPQ \cite{den19} was computed by the GNSS/geoid approach, which is expected to be more accurate than geometric leveling for large distances, as it is not affected by systematic leveling errors.

The ellipsoidal height $h$ from GNSS and the quasigeoid height $\zeta$ are employed to derive the (physically defined) normal height as $H^N = h - \zeta$, which is then converted to a corresponding geopotential value \cite{den17}. 
The (quasi)geoid model EGG2015 \cite{den17} was utilized for this purpose.
The model is based on the remove-compute-restore technique and combines a global long-wavelength spherical-harmonic model from the GOCE mission with high-resolution terrestrial gravity and terrain data. 
The EGG2015 model also includes several gravimetric densification surveys around metrological institutes in Europe, including the PTB and MPQ sites.
Furthermore, error estimates including covariances are available in \cite{den17}.
The GNSS/geoid approach was first used for the computation of the potential difference between the benchmarks near the clock experiments, while the remaining, small height differences between the benchmarks and the clocks were measured by geometric leveling and then converted to a potential difference using $g_\mathrm{MPQ} \approx 9.808 \, \mathrm{m \, s^{-2}}$ and the value at PTB given above, respectively. 

The given uncertainty of $0.30 \, \mathrm{m^2 \, s^{-2}}$ for the geopotential difference is based on the uncertainties (in height) of GNSS positioning (1.0 cm) and the EGG2015 model (1.9 cm), with negligible correlations. 
Furthermore, it should be noted that existing geometric leveling data from the German height network (DHHN92) have also been used to derive the potential difference between the PTB and MPQ clock positions, resulting in a deviation to the recommended value from the GNSS/geoid approach of only $0.3 \, \mathrm{m^2 \, s^{-2}}$ \cite{den17}.

The previously discussed static but spatially variable components of the Earth’s gravity field are the main contributions to the geopotential difference, while time-variable effects (mainly due to solid Earth and ocean tides) are below $0.5 \, \mathrm{m^2 \, s^{-2}}$ (max.~amplitude) for the potential difference between PTB and MPQ.
Furthermore, the time-variable components largely average out over longer recording times and are thus neglected in this analysis. 
The total corresponding height uncertainties are 27~cm for the chronometric value, and 3~cm for the geodetic value.

{\begin{figure}[tb]
\centering
\includegraphics[width=1\linewidth]{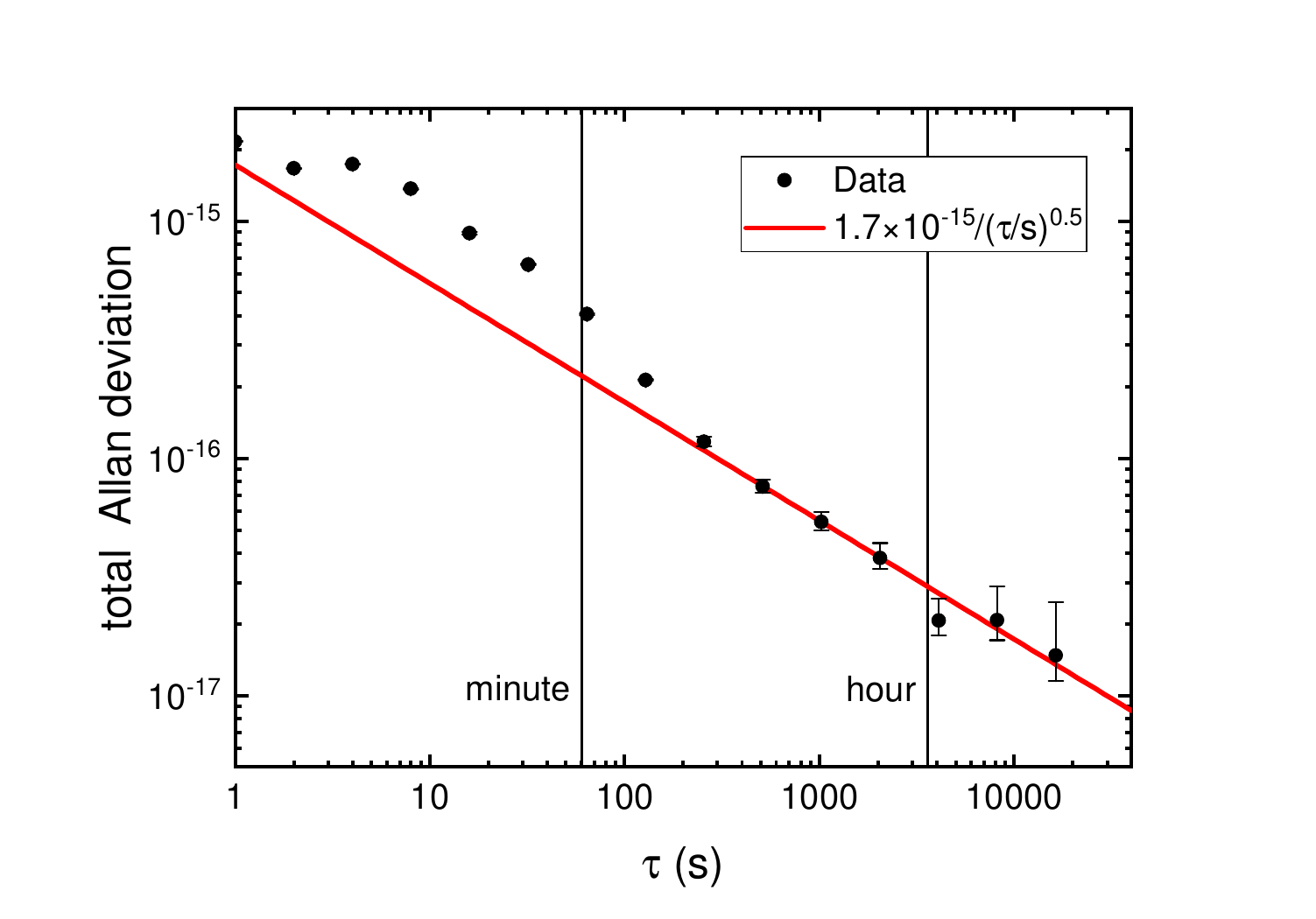}
\caption{(Color online) Total Allan deviation of the concatenated data of the remote clock comparisons as a function of the averaging time $\tau$. The red line indicates the instability and is a $\propto \tau^{-1/2}$ fit to points with $\tau > 200~\mathrm{s}$.}
\label{fig:ADEVs}
\end{figure}}

As shown in Fig.~\ref{fig:ADEVs}, we achieve a height resolution below 30~cm and a corresponding fractional frequency resolution of $3\times 10^{-17}$ within one hour of measurement, and 6~cm within one day.
The instability is limited by the transportable clock Sr2, whereas that of Sr1 is about an order of magnitude lower \cite{sch20d}. 
The main contribution is Dick noise \cite{dic87} of the transportable clock laser \cite{hae20}. Lower transportable clock instabilities can be realized in the near future, for example by using more stable transportable clock lasers \cite{her22} or by novel pre-stabilization methods \cite{kat21}.

In conclusion, we have demonstrated chronometric leveling over a distance of 457~km using a transportable lattice clock, a stationary clock as reference, and an IFL for frequency comparison.
The chronometrically measured geopotential difference is determined with an uncertainty of $2.6 \, \mathrm{m^2 \, s^{-2}}$, corresponding to a height uncertainty of 27~cm.
Further studies will aim to reduce the systematic uncertainty of the transportable clock to the $10^{-18}$ level or below and to reduce its instability and thus the required averaging times. 
This will bring centimeter-level chronometric leveling within reach and could resolve existing discrepancies (at the centimeter to decimeter level) between the established geodetic methods, overcome their limitations, improve continental and global height reference frames, connect island and mainland height systems, connect tide gauges for sea level monitoring, and contribute to gravity field modeling in combination with GNSS positions~\cite{meh18}. As additional benefit, temporal variations of the gravitational field on a daily basis will be ovservable that are currently out of reach for other methods \cite{voi16, tan21b, tak22}.

\begin{acknowledgments}
We thank Thomas Legero and Uwe Sterr for providing a stable laser oscillator and Andreas Koczwara for the development of electronics for the IFL.
We acknowledge support by the Deutsche Forschungsgemeinschaft (DFG, German Research Foundation) under Germany's Excellence Strategy -- EXC-2123 QuantumFrontiers -- Project-ID 3908379.67, SFB~1464 TerraQ -- Project-ID 434617780 -- within projects A04 and A05, and SFB~1128 geoQ within projects A03 and A04.
This work was partially supported by the Max Planck--RIKEN--PTB Center for Time, Constants and Fundamental Symmetries.
\end{acknowledgments}

J.G.\ and I.N.\ contributed equivalently to this work.


\begin{thebibliography}{67}%
\makeatletter
\providecommand \@ifxundefined [1]{%
 \@ifx{#1\undefined}
}%
\providecommand \@ifnum [1]{%
 \ifnum #1\expandafter \@firstoftwo
 \else \expandafter \@secondoftwo
 \fi
}%
\providecommand \@ifx [1]{%
 \ifx #1\expandafter \@firstoftwo
 \else \expandafter \@secondoftwo
 \fi
}%
\providecommand \natexlab [1]{#1}%
\providecommand \enquote  [1]{``#1''}%
\providecommand \bibnamefont  [1]{#1}%
\providecommand \bibfnamefont [1]{#1}%
\providecommand \citenamefont [1]{#1}%
\providecommand \href@noop [0]{\@secondoftwo}%
\providecommand \href [0]{\begingroup \@sanitize@url \@href}%
\providecommand \@href[1]{\@@startlink{#1}\@@href}%
\providecommand \@@href[1]{\endgroup#1\@@endlink}%
\providecommand \@sanitize@url [0]{\catcode `\\12\catcode `\$12\catcode
  `\&12\catcode `\#12\catcode `\^12\catcode `\_12\catcode `\%12\relax}%
\providecommand \@@startlink[1]{}%
\providecommand \@@endlink[0]{}%
\providecommand \url  [0]{\begingroup\@sanitize@url \@url }%
\providecommand \@url [1]{\endgroup\@href {#1}{\urlprefix }}%
\providecommand \urlprefix  [0]{URL }%
\providecommand \Eprint [0]{\href }%
\providecommand \doibase [0]{https://doi.org/}%
\providecommand \selectlanguage [0]{\@gobble}%
\providecommand \bibinfo  [0]{\@secondoftwo}%
\providecommand \bibfield  [0]{\@secondoftwo}%
\providecommand \translation [1]{[#1]}%
\providecommand \BibitemOpen [0]{}%
\providecommand \bibitemStop [0]{}%
\providecommand \bibitemNoStop [0]{.\EOS\space}%
\providecommand \EOS [0]{\spacefactor3000\relax}%
\providecommand \BibitemShut  [1]{\csname bibitem#1\endcsname}%
\let\auto@bib@innerbib\@empty
\bibitem [{\citenamefont {Ushijima}\ \emph {et~al.}(2015)\citenamefont
  {Ushijima}, \citenamefont {Takamoto}, \citenamefont {Das}, \citenamefont
  {Ohkubo},\ and\ \citenamefont {Katori}}]{ush15}%
  \BibitemOpen
  \bibfield  {author} {\bibinfo {author} {\bibfnamefont {I.}~\bibnamefont
  {Ushijima}}, \bibinfo {author} {\bibfnamefont {M.}~\bibnamefont {Takamoto}},
  \bibinfo {author} {\bibfnamefont {M.}~\bibnamefont {Das}}, \bibinfo {author}
  {\bibfnamefont {T.}~\bibnamefont {Ohkubo}},\ and\ \bibinfo {author}
  {\bibfnamefont {H.}~\bibnamefont {Katori}},\ }\bibfield  {title} {\bibinfo
  {title} {Cryogenic optical lattice clocks},\ }\href
  {https://doi.org/10.1038/nphoton.2015.5} {\bibfield  {journal} {\bibinfo
  {journal} {Nat. Photonics}\ }\textbf {\bibinfo {volume} {9}},\ \bibinfo
  {pages} {185} (\bibinfo {year} {2015})}\BibitemShut {NoStop}%
\bibitem [{\citenamefont {Huntemann}\ \emph {et~al.}(2016)\citenamefont
  {Huntemann}, \citenamefont {Sanner}, \citenamefont {Lipphardt}, \citenamefont
  {Tamm},\ and\ \citenamefont {Peik}}]{hun16}%
  \BibitemOpen
  \bibfield  {author} {\bibinfo {author} {\bibfnamefont {N.}~\bibnamefont
  {Huntemann}}, \bibinfo {author} {\bibfnamefont {C.}~\bibnamefont {Sanner}},
  \bibinfo {author} {\bibfnamefont {B.}~\bibnamefont {Lipphardt}}, \bibinfo
  {author} {\bibfnamefont {C.}~\bibnamefont {Tamm}},\ and\ \bibinfo {author}
  {\bibfnamefont {E.}~\bibnamefont {Peik}},\ }\bibfield  {title} {\bibinfo
  {title} {Single-ion atomic clock with $3\times{10}^{-18}$ systematic
  uncertainty},\ }\href {https://doi.org/10.1103/PhysRevLett.116.063001}
  {\bibfield  {journal} {\bibinfo  {journal} {Phys. Rev. Lett.}\ }\textbf
  {\bibinfo {volume} {116}},\ \bibinfo {pages} {063001} (\bibinfo {year}
  {2016})}\BibitemShut {NoStop}%
\bibitem [{\citenamefont {Beloy}\ \emph {et~al.}(2021)\citenamefont {Beloy},
  \citenamefont {Bodine}, \citenamefont {Bothwell}, \citenamefont {Brewer},
  \citenamefont {Bromley}, \citenamefont {Chen}, \citenamefont {Desch{\^e}nes},
  \citenamefont {Diddams}, \citenamefont {Fasano}, \citenamefont {Fortier},
  \citenamefont {Hassan}, \citenamefont {Hume}, \citenamefont {Kedar},
  \citenamefont {Kennedy}, \citenamefont {Khader}, \citenamefont {Koepke},
  \citenamefont {Leibrandt}, \citenamefont {Leopardi}, \citenamefont {Ludlow},
  \citenamefont {McGrew}, \citenamefont {Milner}, \citenamefont {Newbury},
  \citenamefont {Nicolodi}, \citenamefont {Oelker}, \citenamefont {Parker},
  \citenamefont {Robinson}, \citenamefont {Romisch}, \citenamefont
  {Sch\"affer}, \citenamefont {Sherman}, \citenamefont {Sinclair},
  \citenamefont {Sonderhouse}, \citenamefont {Swann}, \citenamefont {Yao},
  \citenamefont {Ye}, \citenamefont {Zhang},\ and\ \citenamefont
  {Collaboration*}}]{bel21}%
  \BibitemOpen
  \bibfield  {author} {\bibinfo {author} {\bibfnamefont {K.}~\bibnamefont
  {Beloy}}, \bibinfo {author} {\bibfnamefont {M.~I.}\ \bibnamefont {Bodine}},
  \bibinfo {author} {\bibfnamefont {T.}~\bibnamefont {Bothwell}}, \bibinfo
  {author} {\bibfnamefont {S.~M.}\ \bibnamefont {Brewer}}, \bibinfo {author}
  {\bibfnamefont {S.~L.}\ \bibnamefont {Bromley}}, \bibinfo {author}
  {\bibfnamefont {J.-S.}\ \bibnamefont {Chen}}, \bibinfo {author}
  {\bibfnamefont {J.-D.}\ \bibnamefont {Desch{\^e}nes}}, \bibinfo {author}
  {\bibfnamefont {S.~A.}\ \bibnamefont {Diddams}}, \bibinfo {author}
  {\bibfnamefont {R.~J.}\ \bibnamefont {Fasano}}, \bibinfo {author}
  {\bibfnamefont {T.~M.}\ \bibnamefont {Fortier}}, \bibinfo {author}
  {\bibfnamefont {Y.~S.}\ \bibnamefont {Hassan}}, \bibinfo {author}
  {\bibfnamefont {D.~B.}\ \bibnamefont {Hume}}, \bibinfo {author}
  {\bibfnamefont {D.}~\bibnamefont {Kedar}}, \bibinfo {author} {\bibfnamefont
  {C.~J.}\ \bibnamefont {Kennedy}}, \bibinfo {author} {\bibfnamefont
  {I.}~\bibnamefont {Khader}}, \bibinfo {author} {\bibfnamefont
  {A.}~\bibnamefont {Koepke}}, \bibinfo {author} {\bibfnamefont {D.~R.}\
  \bibnamefont {Leibrandt}}, \bibinfo {author} {\bibfnamefont {H.}~\bibnamefont
  {Leopardi}}, \bibinfo {author} {\bibfnamefont {A.~D.}\ \bibnamefont
  {Ludlow}}, \bibinfo {author} {\bibfnamefont {W.~F.}\ \bibnamefont {McGrew}},
  \bibinfo {author} {\bibfnamefont {W.~R.}\ \bibnamefont {Milner}}, \bibinfo
  {author} {\bibfnamefont {N.~R.}\ \bibnamefont {Newbury}}, \bibinfo {author}
  {\bibfnamefont {D.}~\bibnamefont {Nicolodi}}, \bibinfo {author}
  {\bibfnamefont {E.}~\bibnamefont {Oelker}}, \bibinfo {author} {\bibfnamefont
  {T.~E.}\ \bibnamefont {Parker}}, \bibinfo {author} {\bibfnamefont {J.~M.}\
  \bibnamefont {Robinson}}, \bibinfo {author} {\bibfnamefont {S.}~\bibnamefont
  {Romisch}}, \bibinfo {author} {\bibfnamefont {S.~A.}\ \bibnamefont
  {Sch\"affer}}, \bibinfo {author} {\bibfnamefont {J.~A.}\ \bibnamefont
  {Sherman}}, \bibinfo {author} {\bibfnamefont {L.~C.}\ \bibnamefont
  {Sinclair}}, \bibinfo {author} {\bibfnamefont {L.}~\bibnamefont
  {Sonderhouse}}, \bibinfo {author} {\bibfnamefont {W.~C.}\ \bibnamefont
  {Swann}}, \bibinfo {author} {\bibfnamefont {J.}~\bibnamefont {Yao}}, \bibinfo
  {author} {\bibfnamefont {J.}~\bibnamefont {Ye}}, and\ \bibinfo {author}
  {\bibfnamefont {X.}~\bibnamefont {Zhang}},\
  }\bibfield  {title} {\bibinfo {title} {Frequency ratio measurements at
  18-digit accuracy using an optical clock network},\ }\href
  {https://doi.org/10.1038/s41586-021-03253-4} {\bibfield  {journal} {\bibinfo
  {journal} {Nature}\ }\textbf {\bibinfo {volume} {591}},\ \bibinfo {pages}
  {564} (\bibinfo {year} {2021})}\BibitemShut {NoStop}%
\bibitem [{\citenamefont {Huang}\ \emph {et~al.}(2022)\citenamefont {Huang},
  \citenamefont {Zhang}, \citenamefont {Zeng}, \citenamefont {Hao},
  \citenamefont {Ma}, \citenamefont {Zhang}, \citenamefont {Guan},
  \citenamefont {Chen}, \citenamefont {Wang},\ and\ \citenamefont
  {Gao}}]{hua22}%
  \BibitemOpen
  \bibfield  {author} {\bibinfo {author} {\bibfnamefont {Y.}~\bibnamefont
  {Huang}}, \bibinfo {author} {\bibfnamefont {B.}~\bibnamefont {Zhang}},
  \bibinfo {author} {\bibfnamefont {M.}~\bibnamefont {Zeng}}, \bibinfo {author}
  {\bibfnamefont {Y.}~\bibnamefont {Hao}}, \bibinfo {author} {\bibfnamefont
  {Z.}~\bibnamefont {Ma}}, \bibinfo {author} {\bibfnamefont {H.}~\bibnamefont
  {Zhang}}, \bibinfo {author} {\bibfnamefont {H.}~\bibnamefont {Guan}},
  \bibinfo {author} {\bibfnamefont {Z.}~\bibnamefont {Chen}}, \bibinfo {author}
  {\bibfnamefont {M.}~\bibnamefont {Wang}},\ and\ \bibinfo {author}
  {\bibfnamefont {K.}~\bibnamefont {Gao}},\ }\bibfield  {title} {\bibinfo
  {title} {Liquid-nitrogen-cooled $\mathrm{Ca}^{+}$ optical clock with
  systematic uncertainty of $3\times 10^{-18}$},\ }\href
  {https://doi.org/10.1103/PhysRevApplied.17.034041} {\bibfield  {journal}
  {\bibinfo  {journal} {Phys. Rev. Appl.}\ }\textbf {\bibinfo {volume} {17}},\
  \bibinfo {pages} {034041} (\bibinfo {year} {2022})}\BibitemShut {NoStop}%
\bibitem [{\citenamefont {Predehl}\ \emph {et~al.}(2012)\citenamefont
  {Predehl}, \citenamefont {Grosche}, \citenamefont {Raupach}, \citenamefont
  {Droste}, \citenamefont {Terra}, \citenamefont {Alnis}, \citenamefont
  {Legero}, \citenamefont {H\"ansch}, \citenamefont {Udem}, \citenamefont
  {Holzwarth},\ and\ \citenamefont {Schnatz}}]{pre12}%
  \BibitemOpen
  \bibfield  {author} {\bibinfo {author} {\bibfnamefont {K.}~\bibnamefont
  {Predehl}}, \bibinfo {author} {\bibfnamefont {G.}~\bibnamefont {Grosche}},
  \bibinfo {author} {\bibfnamefont {S.~M.~F.}\ \bibnamefont {Raupach}},
  \bibinfo {author} {\bibfnamefont {S.}~\bibnamefont {Droste}}, \bibinfo
  {author} {\bibfnamefont {O.}~\bibnamefont {Terra}}, \bibinfo {author}
  {\bibfnamefont {J.}~\bibnamefont {Alnis}}, \bibinfo {author} {\bibfnamefont
  {T.}~\bibnamefont {Legero}}, \bibinfo {author} {\bibfnamefont {T.~W.}\
  \bibnamefont {H\"ansch}}, \bibinfo {author} {\bibfnamefont {T.}~\bibnamefont
  {Udem}}, \bibinfo {author} {\bibfnamefont {R.}~\bibnamefont {Holzwarth}},\
  and\ \bibinfo {author} {\bibfnamefont {H.}~\bibnamefont {Schnatz}},\
  }\bibfield  {title} {\bibinfo {title} {A 920-kilometer optical fiber link for
  frequency metrology at the 19th decimal place},\ }\href
  {https://doi.org/10.1126/science.1218442} {\bibfield  {journal} {\bibinfo
  {journal} {Science}\ }\textbf {\bibinfo {volume} {336}},\ \bibinfo {pages}
  {441} (\bibinfo {year} {2012})}\BibitemShut {NoStop}%
\bibitem [{\citenamefont {Lisdat}\ \emph {et~al.}(2016)\citenamefont {Lisdat},
  \citenamefont {Grosche}, \citenamefont {Quintin}, \citenamefont {Shi},
  \citenamefont {Raupach}, \citenamefont {Grebing}, \citenamefont {Nicolodi},
  \citenamefont {Stefani}, \citenamefont {Al-Masoudi}, \citenamefont
  {D\"orscher}, \citenamefont {H\"afner}, \citenamefont {Robyr}, \citenamefont
  {Chiodo}, \citenamefont {Bilicki}, \citenamefont {Bookjans}, \citenamefont
  {Koczwara}, \citenamefont {Koke}, \citenamefont {Kuhl}, \citenamefont
  {Wiotte}, \citenamefont {Meynadier}, \citenamefont {Camisard}, \citenamefont
  {Abgrall}, \citenamefont {Lours}, \citenamefont {Legero}, \citenamefont
  {Schnatz}, \citenamefont {Sterr}, \citenamefont {Denker}, \citenamefont
  {Chardonnet}, \citenamefont {Le~Coq}, \citenamefont {Santarelli},
  \citenamefont {Amy-Klein}, \citenamefont {Le~Targat}, \citenamefont
  {Lodewyck}, \citenamefont {Lopez},\ and\ \citenamefont {Pottie}}]{lis16}%
  \BibitemOpen
  \bibfield  {author} {\bibinfo {author} {\bibfnamefont {C.}~\bibnamefont
  {Lisdat}}, \bibinfo {author} {\bibfnamefont {G.}~\bibnamefont {Grosche}},
  \bibinfo {author} {\bibfnamefont {N.}~\bibnamefont {Quintin}}, \bibinfo
  {author} {\bibfnamefont {C.}~\bibnamefont {Shi}}, \bibinfo {author}
  {\bibfnamefont {S.}~\bibnamefont {Raupach}}, \bibinfo {author} {\bibfnamefont
  {C.}~\bibnamefont {Grebing}}, \bibinfo {author} {\bibfnamefont
  {D.}~\bibnamefont {Nicolodi}}, \bibinfo {author} {\bibfnamefont
  {F.}~\bibnamefont {Stefani}}, \bibinfo {author} {\bibfnamefont
  {A.}~\bibnamefont {Al-Masoudi}}, \bibinfo {author} {\bibfnamefont
  {S.}~\bibnamefont {D\"orscher}}, \bibinfo {author} {\bibfnamefont
  {S.}~\bibnamefont {H\"afner}}, \bibinfo {author} {\bibfnamefont {J.-L.}\
  \bibnamefont {Robyr}}, \bibinfo {author} {\bibfnamefont {N.}~\bibnamefont
  {Chiodo}}, \bibinfo {author} {\bibfnamefont {S.}~\bibnamefont {Bilicki}},
  \bibinfo {author} {\bibfnamefont {E.}~\bibnamefont {Bookjans}}, \bibinfo
  {author} {\bibfnamefont {A.}~\bibnamefont {Koczwara}}, \bibinfo {author}
  {\bibfnamefont {S.}~\bibnamefont {Koke}}, \bibinfo {author} {\bibfnamefont
  {A.}~\bibnamefont {Kuhl}}, \bibinfo {author} {\bibfnamefont {F.}~\bibnamefont
  {Wiotte}}, \bibinfo {author} {\bibfnamefont {F.}~\bibnamefont {Meynadier}},
  \bibinfo {author} {\bibfnamefont {E.}~\bibnamefont {Camisard}}, \bibinfo
  {author} {\bibfnamefont {M.}~\bibnamefont {Abgrall}}, \bibinfo {author}
  {\bibfnamefont {M.}~\bibnamefont {Lours}}, \bibinfo {author} {\bibfnamefont
  {T.}~\bibnamefont {Legero}}, \bibinfo {author} {\bibfnamefont
  {H.}~\bibnamefont {Schnatz}}, \bibinfo {author} {\bibfnamefont
  {U.}~\bibnamefont {Sterr}}, \bibinfo {author} {\bibfnamefont
  {H.}~\bibnamefont {Denker}}, \bibinfo {author} {\bibfnamefont
  {C.}~\bibnamefont {Chardonnet}}, \bibinfo {author} {\bibfnamefont
  {Y.}~\bibnamefont {Le~Coq}}, \bibinfo {author} {\bibfnamefont
  {G.}~\bibnamefont {Santarelli}}, \bibinfo {author} {\bibfnamefont
  {A.}~\bibnamefont {Amy-Klein}}, \bibinfo {author} {\bibfnamefont
  {R.}~\bibnamefont {Le~Targat}}, \bibinfo {author} {\bibfnamefont
  {J.}~\bibnamefont {Lodewyck}}, \bibinfo {author} {\bibfnamefont
  {O.}~\bibnamefont {Lopez}},\ and\ \bibinfo {author} {\bibfnamefont {P.-E.}\
  \bibnamefont {Pottie}},\ }\bibfield  {title} {\bibinfo {title} {A clock
  network for geodesy and fundamental science},\ }\href
  {https://doi.org/10.1038/ncomms12443} {\bibfield  {journal} {\bibinfo
  {journal} {Nat. Commun.}\ }\textbf {\bibinfo {volume} {7}},\ \bibinfo {pages}
  {12443} (\bibinfo {year} {2016})}\BibitemShut {NoStop}%
\bibitem [{\citenamefont {Cantin}\ \emph {et~al.}(2021)\citenamefont {Cantin},
  \citenamefont {T{\o}nnes}, \citenamefont {Targat}, \citenamefont {Amy-Klein},
  \citenamefont {Lopez},\ and\ \citenamefont {Pottie}}]{can21}%
  \BibitemOpen
  \bibfield  {author} {\bibinfo {author} {\bibfnamefont {E.}~\bibnamefont
  {Cantin}}, \bibinfo {author} {\bibfnamefont {M.}~\bibnamefont {T{\o}nnes}},
  \bibinfo {author} {\bibfnamefont {R.~L.}\ \bibnamefont {Targat}}, \bibinfo
  {author} {\bibfnamefont {A.}~\bibnamefont {Amy-Klein}}, \bibinfo {author}
  {\bibfnamefont {O.}~\bibnamefont {Lopez}},\ and\ \bibinfo {author}
  {\bibfnamefont {P.-E.}\ \bibnamefont {Pottie}},\ }\bibfield  {title}
  {\bibinfo {title} {An accurate and robust metrological network for coherent
  optical frequency dissemination},\ }\href
  {https://doi.org/10.1088/1367-2630/abe79e} {\bibfield  {journal} {\bibinfo
  {journal} {New J. Phys.}\ }\textbf {\bibinfo {volume} {23}},\ \bibinfo
  {pages} {053027} (\bibinfo {year} {2021})}\BibitemShut {NoStop}%
\bibitem [{\citenamefont {Clivati}\ \emph {et~al.}(2020)\citenamefont
  {Clivati}, \citenamefont {Aiello}, \citenamefont {Bianco}, \citenamefont
  {Bortolotti}, \citenamefont {Natale}, \citenamefont {Sarno}, \citenamefont
  {Maddaloni}, \citenamefont {Maccaferri}, \citenamefont {Mura}, \citenamefont
  {Negusini}, \citenamefont {Levi}, \citenamefont {Perini}, \citenamefont
  {Ricci}, \citenamefont {Roma}, \citenamefont {Amato}, \citenamefont
  {de~Cumis}, \citenamefont {Stagni}, \citenamefont {Tuozzi},\ and\
  \citenamefont {Calonico}}]{cli20a}%
  \BibitemOpen
  \bibfield  {author} {\bibinfo {author} {\bibfnamefont {C.}~\bibnamefont
  {Clivati}}, \bibinfo {author} {\bibfnamefont {R.}~\bibnamefont {Aiello}},
  \bibinfo {author} {\bibfnamefont {G.}~\bibnamefont {Bianco}}, \bibinfo
  {author} {\bibfnamefont {C.}~\bibnamefont {Bortolotti}}, \bibinfo {author}
  {\bibfnamefont {P.~D.}\ \bibnamefont {Natale}}, \bibinfo {author}
  {\bibfnamefont {V.~D.}\ \bibnamefont {Sarno}}, \bibinfo {author}
  {\bibfnamefont {P.}~\bibnamefont {Maddaloni}}, \bibinfo {author}
  {\bibfnamefont {G.}~\bibnamefont {Maccaferri}}, \bibinfo {author}
  {\bibfnamefont {A.}~\bibnamefont {Mura}}, \bibinfo {author} {\bibfnamefont
  {M.}~\bibnamefont {Negusini}}, \bibinfo {author} {\bibfnamefont
  {F.}~\bibnamefont {Levi}}, \bibinfo {author} {\bibfnamefont {F.}~\bibnamefont
  {Perini}}, \bibinfo {author} {\bibfnamefont {R.}~\bibnamefont {Ricci}},
  \bibinfo {author} {\bibfnamefont {M.}~\bibnamefont {Roma}}, \bibinfo {author}
  {\bibfnamefont {L.~S.}\ \bibnamefont {Amato}}, \bibinfo {author}
  {\bibfnamefont {M.~S.}\ \bibnamefont {de~Cumis}}, \bibinfo {author}
  {\bibfnamefont {M.}~\bibnamefont {Stagni}}, \bibinfo {author} {\bibfnamefont
  {A.}~\bibnamefont {Tuozzi}},\ and\ \bibinfo {author} {\bibfnamefont
  {D.}~\bibnamefont {Calonico}},\ }\bibfield  {title} {\bibinfo {title}
  {Common-clock very long baseline interferometry using a coherent optical
  fiber link},\ }\href {https://doi.org/10.1364/OPTICA.393356} {\bibfield
  {journal} {\bibinfo  {journal} {Optica}\ }\textbf {\bibinfo {volume} {7}},\
  \bibinfo {pages} {1031} (\bibinfo {year} {2020})}\BibitemShut {NoStop}%
\bibitem [{\citenamefont {Akatsuka}\ \emph {et~al.}(2020)\citenamefont
  {Akatsuka}, \citenamefont {Goh}, \citenamefont {Imai}, \citenamefont {Oguri},
  \citenamefont {Ishizawa}, \citenamefont {Ushijima}, \citenamefont {Ushijima},
  \citenamefont {Ohmae}, \citenamefont {Ohmae}, \citenamefont {Takamoto},
  \citenamefont {Takamoto}, \citenamefont {Katori}, \citenamefont {Katori},
  \citenamefont {Katori}, \citenamefont {Hashimoto}, \citenamefont {Gotoh},\
  and\ \citenamefont {Sogawa}}]{aka20}%
  \BibitemOpen
  \bibfield  {author} {\bibinfo {author} {\bibfnamefont {T.}~\bibnamefont
  {Akatsuka}}, \bibinfo {author} {\bibfnamefont {T.}~\bibnamefont {Goh}},
  \bibinfo {author} {\bibfnamefont {H.}~\bibnamefont {Imai}}, \bibinfo {author}
  {\bibfnamefont {K.}~\bibnamefont {Oguri}}, \bibinfo {author} {\bibfnamefont
  {A.}~\bibnamefont {Ishizawa}}, \bibinfo {author} {\bibfnamefont
  {I.}~\bibnamefont {Ushijima}}, \bibinfo {author} {\bibfnamefont
  {I.}~\bibnamefont {Ushijima}}, \bibinfo {author} {\bibfnamefont
  {N.}~\bibnamefont {Ohmae}}, \bibinfo {author} {\bibfnamefont
  {N.}~\bibnamefont {Ohmae}}, \bibinfo {author} {\bibfnamefont
  {M.}~\bibnamefont {Takamoto}}, \bibinfo {author} {\bibfnamefont
  {M.}~\bibnamefont {Takamoto}}, \bibinfo {author} {\bibfnamefont
  {H.}~\bibnamefont {Katori}}, \bibinfo {author} {\bibfnamefont
  {H.}~\bibnamefont {Katori}}, \bibinfo {author} {\bibfnamefont
  {H.}~\bibnamefont {Katori}}, \bibinfo {author} {\bibfnamefont
  {T.}~\bibnamefont {Hashimoto}}, \bibinfo {author} {\bibfnamefont
  {H.}~\bibnamefont {Gotoh}},\ and\ \bibinfo {author} {\bibfnamefont
  {T.}~\bibnamefont {Sogawa}},\ }\bibfield  {title} {\bibinfo {title} {Optical
  frequency distribution using laser repeater stations with planar lightwave
  circuits},\ }\href {https://doi.org/10.1364/OE.383526} {\bibfield  {journal}
  {\bibinfo  {journal} {Opt. Express}\ }\textbf {\bibinfo {volume} {28}},\
  \bibinfo {pages} {9186} (\bibinfo {year} {2020})}\BibitemShut {NoStop}%
\bibitem [{\citenamefont {Schioppo}\ \emph {et~al.}(2022)\citenamefont
  {Schioppo}, \citenamefont {Kronj\"ager}, \citenamefont {Silva}, \citenamefont
  {Ilieva}, \citenamefont {Paterson}, \citenamefont {Baynham}, \citenamefont
  {Bowden}, \citenamefont {Hill}, \citenamefont {Hobson}, \citenamefont
  {Vianello}, \citenamefont {Dovale-\'Alvarez}, \citenamefont {Williams},
  \citenamefont {Marra}, \citenamefont {Margolis}, \citenamefont {Amy-Klein},
  \citenamefont {Lopez}, \citenamefont {Cantin}, \citenamefont
  {\'Alvarez-Mart\'inez}, \citenamefont {Le~Targat}, \citenamefont {Pottie},
  \citenamefont {Quintin}, \citenamefont {Legero}, \citenamefont {H\"afner},
  \citenamefont {Sterr}, \citenamefont {Schwarz}, \citenamefont {D\"orscher},
  \citenamefont {Lisdat}, \citenamefont {Koke}, \citenamefont {Kuhl},
  \citenamefont {Waterholter}, \citenamefont {Benkler},\ and\ \citenamefont
  {Grosche}}]{sch22a}%
  \BibitemOpen
  \bibfield  {author} {\bibinfo {author} {\bibfnamefont {M.}~\bibnamefont
  {Schioppo}}, \bibinfo {author} {\bibfnamefont {J.}~\bibnamefont
  {Kronj\"ager}}, \bibinfo {author} {\bibfnamefont {A.}~\bibnamefont {Silva}},
  \bibinfo {author} {\bibfnamefont {R.}~\bibnamefont {Ilieva}}, \bibinfo
  {author} {\bibfnamefont {J.~W.}\ \bibnamefont {Paterson}}, \bibinfo {author}
  {\bibfnamefont {C.~F.~A.}\ \bibnamefont {Baynham}}, \bibinfo {author}
  {\bibfnamefont {W.}~\bibnamefont {Bowden}}, \bibinfo {author} {\bibfnamefont
  {I.~R.}\ \bibnamefont {Hill}}, \bibinfo {author} {\bibfnamefont
  {R.}~\bibnamefont {Hobson}}, \bibinfo {author} {\bibfnamefont
  {A.}~\bibnamefont {Vianello}}, \bibinfo {author} {\bibfnamefont
  {M.}~\bibnamefont {Dovale-\'Alvarez}}, \bibinfo {author} {\bibfnamefont
  {R.~A.}\ \bibnamefont {Williams}}, \bibinfo {author} {\bibfnamefont
  {G.}~\bibnamefont {Marra}}, \bibinfo {author} {\bibfnamefont {H.~S.}\
  \bibnamefont {Margolis}}, \bibinfo {author} {\bibfnamefont {A.}~\bibnamefont
  {Amy-Klein}}, \bibinfo {author} {\bibfnamefont {O.}~\bibnamefont {Lopez}},
  \bibinfo {author} {\bibfnamefont {E.}~\bibnamefont {Cantin}}, \bibinfo
  {author} {\bibfnamefont {H.}~\bibnamefont {\'Alvarez-Mart\'inez}}, \bibinfo
  {author} {\bibfnamefont {R.}~\bibnamefont {Le~Targat}}, \bibinfo {author}
  {\bibfnamefont {P.~E.}\ \bibnamefont {Pottie}}, \bibinfo {author}
  {\bibfnamefont {N.}~\bibnamefont {Quintin}}, \bibinfo {author} {\bibfnamefont
  {T.}~\bibnamefont {Legero}}, \bibinfo {author} {\bibfnamefont
  {S.}~\bibnamefont {H\"afner}}, \bibinfo {author} {\bibfnamefont
  {U.}~\bibnamefont {Sterr}}, \bibinfo {author} {\bibfnamefont
  {R.}~\bibnamefont {Schwarz}}, \bibinfo {author} {\bibfnamefont
  {S.}~\bibnamefont {D\"orscher}}, \bibinfo {author} {\bibfnamefont
  {C.}~\bibnamefont {Lisdat}}, \bibinfo {author} {\bibfnamefont
  {S.}~\bibnamefont {Koke}}, \bibinfo {author} {\bibfnamefont {A.}~\bibnamefont
  {Kuhl}}, \bibinfo {author} {\bibfnamefont {T.}~\bibnamefont {Waterholter}},
  \bibinfo {author} {\bibfnamefont {E.}~\bibnamefont {Benkler}},\ and\ \bibinfo
  {author} {\bibfnamefont {G.}~\bibnamefont {Grosche}},\ }\bibfield  {title}
  {\bibinfo {title} {Comparing ultrastable lasers at $7\times 10^{-17}$
  fractional frequency instability through a 2220~km optical fibre network},\
  }\href {https://doi.org/10.1038/s41467-021-27884-3} {\bibfield  {journal}
  {\bibinfo  {journal} {Nat. Commun.}\ }\textbf {\bibinfo {volume} {13}},\
  \bibinfo {pages} {212} (\bibinfo {year} {2022})}\BibitemShut {NoStop}%
\bibitem [{\citenamefont {Vermeer}(1983)}]{ver83}%
  \BibitemOpen
  \bibfield  {author} {\bibinfo {author} {\bibfnamefont {M.}~\bibnamefont
  {Vermeer}},\ }\href {http://books.google.de/books?id=-iaUAAAACAAJ} {\emph
  {\bibinfo {title} {Chronometric Levelling}}},\ \bibinfo {series} {Reports of
  the Finnish Geodetic Institute}\ No.\ \bibinfo {number} {83:2}\ (\bibinfo
  {publisher} {Geodeettinen Laitos, Geodetiska Institutet},\
  \bibinfo{location} {Helsinki,\ Finland},\ \bibinfo {year}
  {1983})\BibitemShut {NoStop}%
\bibitem [{\citenamefont {Bjerhammar}(1985)}]{bje85}%
  \BibitemOpen
  \bibfield  {author} {\bibinfo {author} {\bibfnamefont {A.}~\bibnamefont
  {Bjerhammar}},\ }\bibfield  {title} {\bibinfo {title} {On a relativistic
  geodesy},\ }\href {https://doi.org/10.1007/BF02520327} {\bibfield  {journal}
  {\bibinfo  {journal} {Bulletin G\'eod\'esique}\ }\textbf {\bibinfo {volume}
  {59}},\ \bibinfo {pages} {207} (\bibinfo {year} {1985})}\BibitemShut
  {NoStop}%
\bibitem [{\citenamefont {Mehlst{\"a}ubler}\ \emph {et~al.}(2018)\citenamefont
  {Mehlst{\"a}ubler}, \citenamefont {Grosche}, \citenamefont {Lisdat},
  \citenamefont {Schmidt},\ and\ \citenamefont {Denker}}]{meh18}%
  \BibitemOpen
  \bibfield  {author} {\bibinfo {author} {\bibfnamefont {T.}~\bibnamefont
  {Mehlst{\"a}ubler}}, \bibinfo {author} {\bibfnamefont {G.}~\bibnamefont
  {Grosche}}, \bibinfo {author} {\bibfnamefont {C.}~\bibnamefont {Lisdat}},
  \bibinfo {author} {\bibfnamefont {P.}~\bibnamefont {Schmidt}},\ and\ \bibinfo
  {author} {\bibfnamefont {H.}~\bibnamefont {Denker}},\ }\bibfield  {title}
  {\bibinfo {title} {Atomic clocks for geodesy},\ }\href
  {https://doi.org/10.1088/1361-6633/aab409} {\bibfield  {journal} {\bibinfo
  {journal} {Rep. Prog. Phys.}\ }\textbf {\bibinfo {volume} {81}},\ \bibinfo
  {pages} {064401} (\bibinfo {year} {2018})}\BibitemShut {NoStop}%
\bibitem [{\citenamefont {Chou}\ \emph {et~al.}(2010)\citenamefont {Chou},
  \citenamefont {Hume}, \citenamefont {Rosenband},\ and\ \citenamefont
  {Wineland}}]{cho10a}%
  \BibitemOpen
  \bibfield  {author} {\bibinfo {author} {\bibfnamefont {C.~W.}\ \bibnamefont
  {Chou}}, \bibinfo {author} {\bibfnamefont {D.~B.}\ \bibnamefont {Hume}},
  \bibinfo {author} {\bibfnamefont {T.}~\bibnamefont {Rosenband}},\ and\
  \bibinfo {author} {\bibfnamefont {D.~J.}\ \bibnamefont {Wineland}},\
  }\bibfield  {title} {\bibinfo {title} {Optical clocks and relativity},\
  }\href {https://doi.org/10.1126/science.1192720} {\bibfield  {journal}
  {\bibinfo  {journal} {Science}\ }\textbf {\bibinfo {volume} {329}},\ \bibinfo
  {pages} {1630 } (\bibinfo {year} {2010})}\BibitemShut {NoStop}%
\bibitem [{\citenamefont {Takamoto}\ \emph {et~al.}(2020)\citenamefont
  {Takamoto}, \citenamefont {Ushijima}, \citenamefont {Ohmae}, \citenamefont
  {Yahagi}, \citenamefont {Kokado}, \citenamefont {Shinkai},\ and\
  \citenamefont {Katori}}]{tak20}%
  \BibitemOpen
  \bibfield  {author} {\bibinfo {author} {\bibfnamefont {M.}~\bibnamefont
  {Takamoto}}, \bibinfo {author} {\bibfnamefont {I.}~\bibnamefont {Ushijima}},
  \bibinfo {author} {\bibfnamefont {N.}~\bibnamefont {Ohmae}}, \bibinfo
  {author} {\bibfnamefont {T.}~\bibnamefont {Yahagi}}, \bibinfo {author}
  {\bibfnamefont {K.}~\bibnamefont {Kokado}}, \bibinfo {author} {\bibfnamefont
  {H.}~\bibnamefont {Shinkai}},\ and\ \bibinfo {author} {\bibfnamefont
  {H.}~\bibnamefont {Katori}},\ }\bibfield  {title} {\bibinfo {title} {Test of
  general relativity by a pair of transportable optical lattice clocks},\
  }\href {https://doi.org/10.1038/s41566-020-0619-8} {\bibfield  {journal}
  {\bibinfo  {journal} {Nat. Photonics}\ }\textbf {\bibinfo {volume} {14}},\
  \bibinfo {pages} {411} (\bibinfo {year} {2020})}\BibitemShut {NoStop}%
\bibitem [{\citenamefont {Huang}\ \emph {et~al.}(2020)\citenamefont {Huang},
  \citenamefont {Zhang}, \citenamefont {Zhang}, \citenamefont {Hao},
  \citenamefont {Guan}, \citenamefont {Zeng}, \citenamefont {Chen},
  \citenamefont {Lin}, \citenamefont {Wang}, \citenamefont {Cao}, \citenamefont
  {Liang}, \citenamefont {Fang}, \citenamefont {Fang}, \citenamefont {Li},\
  and\ \citenamefont {Gao}}]{hua20}%
  \BibitemOpen
  \bibfield  {author} {\bibinfo {author} {\bibfnamefont {Y.}~\bibnamefont
  {Huang}}, \bibinfo {author} {\bibfnamefont {H.}~\bibnamefont {Zhang}},
  \bibinfo {author} {\bibfnamefont {B.}~\bibnamefont {Zhang}}, \bibinfo
  {author} {\bibfnamefont {Y.}~\bibnamefont {Hao}}, \bibinfo {author}
  {\bibfnamefont {H.}~\bibnamefont {Guan}}, \bibinfo {author} {\bibfnamefont
  {M.}~\bibnamefont {Zeng}}, \bibinfo {author} {\bibfnamefont {Q.}~\bibnamefont
  {Chen}}, \bibinfo {author} {\bibfnamefont {Y.}~\bibnamefont {Lin}}, \bibinfo
  {author} {\bibfnamefont {Y.}~\bibnamefont {Wang}}, \bibinfo {author}
  {\bibfnamefont {S.}~\bibnamefont {Cao}}, \bibinfo {author} {\bibfnamefont
  {K.}~\bibnamefont {Liang}}, \bibinfo {author} {\bibfnamefont
  {F.}~\bibnamefont {Fang}}, \bibinfo {author} {\bibfnamefont {Z.}~\bibnamefont
  {Fang}}, \bibinfo {author} {\bibfnamefont {T.}~\bibnamefont {Li}},\ and\
  \bibinfo {author} {\bibfnamefont {K.}~\bibnamefont {Gao}},\ }\bibfield
  {title} {\bibinfo {title} {Geopotential measurement with a robust,
  transportable {Ca$^+$} optical clock},\ }\href
  {https://doi.org/10.1103/PhysRevA.102.050802} {\bibfield  {journal} {\bibinfo
   {journal} {Phys. Rev. A}\ }\textbf {\bibinfo {volume} {102}},\ \bibinfo
  {pages} {050802(R)} (\bibinfo {year} {2020})}\BibitemShut {NoStop}%
\bibitem [{\citenamefont {Liu}\ \emph {et~al.}(2023)\citenamefont {Liu},
  \citenamefont {Cao}, \citenamefont {Yuan}, \citenamefont {Cui}, \citenamefont
  {Yuan}, \citenamefont {Zhang}, \citenamefont {Chao}, \citenamefont {Shu},\
  and\ \citenamefont {Huang}}]{liu23}%
  \BibitemOpen
  \bibfield  {author} {\bibinfo {author} {\bibfnamefont {D.-X.}\ \bibnamefont
  {Liu}}, \bibinfo {author} {\bibfnamefont {J.}~\bibnamefont {Cao}}, \bibinfo
  {author} {\bibfnamefont {J.-B.}\ \bibnamefont {Yuan}}, \bibinfo {author}
  {\bibfnamefont {K.-F.}\ \bibnamefont {Cui}}, \bibinfo {author} {\bibfnamefont
  {Y.}~\bibnamefont {Yuan}}, \bibinfo {author} {\bibfnamefont {P.}~\bibnamefont
  {Zhang}}, \bibinfo {author} {\bibfnamefont {S.-J.}\ \bibnamefont {Chao}},
  \bibinfo {author} {\bibfnamefont {H.-L.}\ \bibnamefont {Shu}},\ and\ \bibinfo
  {author} {\bibfnamefont {X.-R.}\ \bibnamefont {Huang}},\ }\bibfield  {title}
  {\bibinfo {title} {Laboratory demonstration of geopotential measurement using
  transportable optical clocks},\ }\href
  {https://doi.org/10.1088/1674-1056/ac6337} {\bibfield  {journal} {\bibinfo
  {journal} {Chin. Phys. B}\ }\textbf {\bibinfo {volume} {32}},\ \bibinfo
  {pages} {010601} (\bibinfo {year} {2023})}\BibitemShut {NoStop}%
\bibitem [{\citenamefont {Bothwell}\ \emph {et~al.}(2022)\citenamefont
  {Bothwell}, \citenamefont {Kennedy}, \citenamefont {Aeppli}, \citenamefont
  {Kedar}, \citenamefont {Robinson}, \citenamefont {Oelker}, \citenamefont
  {Staron},\ and\ \citenamefont {Ye}}]{bot22}%
  \BibitemOpen
  \bibfield  {author} {\bibinfo {author} {\bibfnamefont {T.}~\bibnamefont
  {Bothwell}}, \bibinfo {author} {\bibfnamefont {C.~J.}\ \bibnamefont
  {Kennedy}}, \bibinfo {author} {\bibfnamefont {A.}~\bibnamefont {Aeppli}},
  \bibinfo {author} {\bibfnamefont {D.}~\bibnamefont {Kedar}}, \bibinfo
  {author} {\bibfnamefont {J.~M.}\ \bibnamefont {Robinson}}, \bibinfo {author}
  {\bibfnamefont {E.}~\bibnamefont {Oelker}}, \bibinfo {author} {\bibfnamefont
  {A.}~\bibnamefont {Staron}},\ and\ \bibinfo {author} {\bibfnamefont
  {J.}~\bibnamefont {Ye}},\ }\bibfield  {title} {\bibinfo {title} {Resolving
  the gravitational redshift across a millimetre-scale atomic sample},\ }\href
  {https://doi.org/10.1038/s41586-021-04349-7} {\bibfield  {journal} {\bibinfo
  {journal} {Nature}\ }\textbf {\bibinfo {volume} {602}},\ \bibinfo {pages}
  {420} (\bibinfo {year} {2022})}\BibitemShut {NoStop}%
\bibitem [{\citenamefont {Zheng}\ \emph {et~al.}(2023)\citenamefont {Zheng},
  \citenamefont {Dolde}, \citenamefont {Cambria}, \citenamefont {Lim},\ and\
  \citenamefont {Kolkowitz}}]{zhe23b}%
  \BibitemOpen
  \bibfield  {author} {\bibinfo {author} {\bibfnamefont {X.}~\bibnamefont
  {Zheng}}, \bibinfo {author} {\bibfnamefont {J.}~\bibnamefont {Dolde}},
  \bibinfo {author} {\bibfnamefont {M.~C.}\ \bibnamefont {Cambria}}, \bibinfo
  {author} {\bibfnamefont {H.~M.}\ \bibnamefont {Lim}},\ and\ \bibinfo {author}
  {\bibfnamefont {S.}~\bibnamefont {Kolkowitz}},\ }\bibfield  {title} {\bibinfo
  {title} {A lab-based test of the gravitational redshift with a miniature
  clock network},\ }\href {https://doi.org/10.1038/s41467-023-40629-8}
  {\bibfield  {journal} {\bibinfo  {journal} {Nat. Commun.}\ }\textbf {\bibinfo
  {volume} {14}},\ \bibinfo {pages} {4886} (\bibinfo {year}
  {2023})}\BibitemShut {NoStop}%
\bibitem [{\citenamefont {Takano}\ \emph {et~al.}(2016)\citenamefont {Takano},
  \citenamefont {Takamoto}, \citenamefont {Ushijima}, \citenamefont {Ohmae},
  \citenamefont {Akatsuka}, \citenamefont {Yamaguchi}, \citenamefont
  {Kuroishi}, \citenamefont {Munekane}, \citenamefont {Miyahara},\ and\
  \citenamefont {Katori}}]{tak16}%
  \BibitemOpen
  \bibfield  {author} {\bibinfo {author} {\bibfnamefont {T.}~\bibnamefont
  {Takano}}, \bibinfo {author} {\bibfnamefont {M.}~\bibnamefont {Takamoto}},
  \bibinfo {author} {\bibfnamefont {I.}~\bibnamefont {Ushijima}}, \bibinfo
  {author} {\bibfnamefont {N.}~\bibnamefont {Ohmae}}, \bibinfo {author}
  {\bibfnamefont {T.}~\bibnamefont {Akatsuka}}, \bibinfo {author}
  {\bibfnamefont {A.}~\bibnamefont {Yamaguchi}}, \bibinfo {author}
  {\bibfnamefont {Y.}~\bibnamefont {Kuroishi}}, \bibinfo {author}
  {\bibfnamefont {H.}~\bibnamefont {Munekane}}, \bibinfo {author}
  {\bibfnamefont {B.}~\bibnamefont {Miyahara}},\ and\ \bibinfo {author}
  {\bibfnamefont {H.}~\bibnamefont {Katori}},\ }\bibfield  {title} {\bibinfo
  {title} {Geopotential measurements with synchronously linked optical lattice
  clocks},\ }\href {https://doi.org/10.1038/nphoton.2016.159} {\bibfield
  {journal} {\bibinfo  {journal} {Nat. Photonics}\ }\textbf {\bibinfo {volume}
  {10}},\ \bibinfo {pages} {662} (\bibinfo {year} {2016})}\BibitemShut
  {NoStop}%
\bibitem [{\citenamefont {Grotti}\ \emph {et~al.}(2018)\citenamefont {Grotti},
  \citenamefont {Koller}, \citenamefont {Vogt}, \citenamefont {H\"afner},
  \citenamefont {Sterr}, \citenamefont {Lisdat}, \citenamefont {Denker},
  \citenamefont {Voigt}, \citenamefont {Timmen}, \citenamefont {Rolland},
  \citenamefont {Baynes}, \citenamefont {Margolis}, \citenamefont {Zampaolo},
  \citenamefont {Thoumany}, \citenamefont {Pizzocaro}, \citenamefont {Rauf},
  \citenamefont {Bregolin}, \citenamefont {Tampellini}, \citenamefont
  {Barbieri}, \citenamefont {Zucco}, \citenamefont {Costanzo}, \citenamefont
  {Clivati}, \citenamefont {Levi},\ and\ \citenamefont {Calonico}}]{gro18a}%
  \BibitemOpen
  \bibfield  {author} {\bibinfo {author} {\bibfnamefont {J.}~\bibnamefont
  {Grotti}}, \bibinfo {author} {\bibfnamefont {S.}~\bibnamefont {Koller}},
  \bibinfo {author} {\bibfnamefont {S.}~\bibnamefont {Vogt}}, \bibinfo {author}
  {\bibfnamefont {S.}~\bibnamefont {H\"afner}}, \bibinfo {author}
  {\bibfnamefont {U.}~\bibnamefont {Sterr}}, \bibinfo {author} {\bibfnamefont
  {C.}~\bibnamefont {Lisdat}}, \bibinfo {author} {\bibfnamefont
  {H.}~\bibnamefont {Denker}}, \bibinfo {author} {\bibfnamefont
  {C.}~\bibnamefont {Voigt}}, \bibinfo {author} {\bibfnamefont
  {L.}~\bibnamefont {Timmen}}, \bibinfo {author} {\bibfnamefont
  {A.}~\bibnamefont {Rolland}}, \bibinfo {author} {\bibfnamefont {F.~N.}\
  \bibnamefont {Baynes}}, \bibinfo {author} {\bibfnamefont {H.~S.}\
  \bibnamefont {Margolis}}, \bibinfo {author} {\bibfnamefont {M.}~\bibnamefont
  {Zampaolo}}, \bibinfo {author} {\bibfnamefont {P.}~\bibnamefont {Thoumany}},
  \bibinfo {author} {\bibfnamefont {M.}~\bibnamefont {Pizzocaro}}, \bibinfo
  {author} {\bibfnamefont {B.}~\bibnamefont {Rauf}}, \bibinfo {author}
  {\bibfnamefont {F.}~\bibnamefont {Bregolin}}, \bibinfo {author}
  {\bibfnamefont {A.}~\bibnamefont {Tampellini}}, \bibinfo {author}
  {\bibfnamefont {P.}~\bibnamefont {Barbieri}}, \bibinfo {author}
  {\bibfnamefont {M.}~\bibnamefont {Zucco}}, \bibinfo {author} {\bibfnamefont
  {G.~A.}\ \bibnamefont {Costanzo}}, \bibinfo {author} {\bibfnamefont
  {C.}~\bibnamefont {Clivati}}, \bibinfo {author} {\bibfnamefont
  {F.}~\bibnamefont {Levi}},\ and\ \bibinfo {author} {\bibfnamefont
  {D.}~\bibnamefont {Calonico}},\ }\bibfield  {title} {\bibinfo {title}
  {Geodesy and metrology with a transportable optical clock},\ }\href
  {https://doi.org/10.1038/s41567-017-0042-3} {\bibfield  {journal} {\bibinfo
  {journal} {Nat. Phys.}\ }\textbf {\bibinfo {volume} {14}},\ \bibinfo {pages}
  {437} (\bibinfo {year} {2018})}\BibitemShut {NoStop}%
\bibitem [{\citenamefont {S{\'a}nchez}\ \emph {et~al.}(2021)\citenamefont
  {S{\'a}nchez}, \citenamefont {{\AA}gren}, \citenamefont {Huang},
  \citenamefont {Wang}, \citenamefont {M\"akinen}, \citenamefont {Pail},
  \citenamefont {Barzaghi}, \citenamefont {Vergos}, \citenamefont {Ahlgren},\
  and\ \citenamefont {Liu}}]{san21a}%
  \BibitemOpen
  \bibfield  {author} {\bibinfo {author} {\bibfnamefont {L.}~\bibnamefont
  {S{\'a}nchez}}, \bibinfo {author} {\bibfnamefont {J.}~\bibnamefont
  {{\AA}gren}}, \bibinfo {author} {\bibfnamefont {J.}~\bibnamefont {Huang}},
  \bibinfo {author} {\bibfnamefont {Y.~M.}\ \bibnamefont {Wang}}, \bibinfo
  {author} {\bibfnamefont {J.}~\bibnamefont {M\"akinen}}, \bibinfo {author}
  {\bibfnamefont {R.}~\bibnamefont {Pail}}, \bibinfo {author} {\bibfnamefont
  {R.}~\bibnamefont {Barzaghi}}, \bibinfo {author} {\bibfnamefont {G.~S.}\
  \bibnamefont {Vergos}}, \bibinfo {author} {\bibfnamefont {K.}~\bibnamefont
  {Ahlgren}},\ and\ \bibinfo {author} {\bibfnamefont {Q.}~\bibnamefont {Liu}},\
  }\bibfield  {title} {\bibinfo {title} {Strategy for the realisation of the
  international height reference system ({IHRS})},\ }\href
  {https://doi.org/10.1007/s00190-021-01481-0} {\bibfield  {journal} {\bibinfo
  {journal} {J. Geod.}\ }\textbf {\bibinfo {volume} {95}},\ \bibinfo {pages}
  {33} (\bibinfo {year} {2021})}\BibitemShut {NoStop}%
\bibitem [{\citenamefont {Denker}\ \emph {et~al.}(2018)\citenamefont {Denker},
  \citenamefont {Timmen}, \citenamefont {Voigt}, \citenamefont {Weyers},
  \citenamefont {Peik}, \citenamefont {Margolis}, \citenamefont {Delva},
  \citenamefont {Wolf},\ and\ \citenamefont {Petit}}]{den17}%
  \BibitemOpen
  \bibfield  {author} {\bibinfo {author} {\bibfnamefont {H.}~\bibnamefont
  {Denker}}, \bibinfo {author} {\bibfnamefont {L.}~\bibnamefont {Timmen}},
  \bibinfo {author} {\bibfnamefont {C.}~\bibnamefont {Voigt}}, \bibinfo
  {author} {\bibfnamefont {S.}~\bibnamefont {Weyers}}, \bibinfo {author}
  {\bibfnamefont {E.}~\bibnamefont {Peik}}, \bibinfo {author} {\bibfnamefont
  {H.~S.}\ \bibnamefont {Margolis}}, \bibinfo {author} {\bibfnamefont
  {P.}~\bibnamefont {Delva}}, \bibinfo {author} {\bibfnamefont
  {P.}~\bibnamefont {Wolf}},\ and\ \bibinfo {author} {\bibfnamefont
  {G.}~\bibnamefont {Petit}},\ }\bibfield  {title} {\bibinfo {title} {Geodetic
  methods to determine the relativistic redshift at the level of $10^{-18}$ in
  the context of international timescales: a review and practical results},\
  }\href {https://doi.org/10.1007/s00190-017-1075-1} {\bibfield  {journal}
  {\bibinfo  {journal} {J. Geod.}\ }\textbf {\bibinfo {volume} {5}},\ \bibinfo
  {pages} {487} (\bibinfo {year} {2018})}\BibitemShut {NoStop}%
\bibitem [{\citenamefont {Koller}\ \emph {et~al.}(2017)\citenamefont {Koller},
  \citenamefont {Grotti}, \citenamefont {Vogt}, \citenamefont {Al-Masoudi},
  \citenamefont {D\"orscher}, \citenamefont {H\"afner}, \citenamefont {Sterr},\
  and\ \citenamefont {Lisdat}}]{kol17}%
  \BibitemOpen
  \bibfield  {author} {\bibinfo {author} {\bibfnamefont {S.~B.}\ \bibnamefont
  {Koller}}, \bibinfo {author} {\bibfnamefont {J.}~\bibnamefont {Grotti}},
  \bibinfo {author} {\bibfnamefont {S.}~\bibnamefont {Vogt}}, \bibinfo {author}
  {\bibfnamefont {A.}~\bibnamefont {Al-Masoudi}}, \bibinfo {author}
  {\bibfnamefont {S.}~\bibnamefont {D\"orscher}}, \bibinfo {author}
  {\bibfnamefont {S.}~\bibnamefont {H\"afner}}, \bibinfo {author}
  {\bibfnamefont {U.}~\bibnamefont {Sterr}},\ and\ \bibinfo {author}
  {\bibfnamefont {C.}~\bibnamefont {Lisdat}},\ }\bibfield  {title} {\bibinfo
  {title} {Transportable optical lattice clock with $7 \times 10^{-17}$
  uncertainty},\ }\href {https://doi.org/10.1103/PhysRevLett.118.073601}
  {\bibfield  {journal} {\bibinfo  {journal} {Phys. Rev. Lett.}\ }\textbf
  {\bibinfo {volume} {118}},\ \bibinfo {pages} {073601} (\bibinfo {year}
  {2017})}\BibitemShut {NoStop}%
\bibitem [{\citenamefont {Origlia}\ \emph {et~al.}(2018)\citenamefont
  {Origlia}, \citenamefont {Pramod}, \citenamefont {Schiller}, \citenamefont
  {Singh}, \citenamefont {Bongs}, \citenamefont {Schwarz}, \citenamefont
  {Al-Masoudi}, \citenamefont {D\"orscher}, \citenamefont {Herbers},
  \citenamefont {H\"afner}, \citenamefont {Sterr},\ and\ \citenamefont
  {Lisdat}}]{ori18}%
  \BibitemOpen
  \bibfield  {author} {\bibinfo {author} {\bibfnamefont {S.}~\bibnamefont
  {Origlia}}, \bibinfo {author} {\bibfnamefont {M.~S.}\ \bibnamefont {Pramod}},
  \bibinfo {author} {\bibfnamefont {S.}~\bibnamefont {Schiller}}, \bibinfo
  {author} {\bibfnamefont {Y.}~\bibnamefont {Singh}}, \bibinfo {author}
  {\bibfnamefont {K.}~\bibnamefont {Bongs}}, \bibinfo {author} {\bibfnamefont
  {R.}~\bibnamefont {Schwarz}}, \bibinfo {author} {\bibfnamefont
  {A.}~\bibnamefont {Al-Masoudi}}, \bibinfo {author} {\bibfnamefont
  {S.}~\bibnamefont {D\"orscher}}, \bibinfo {author} {\bibfnamefont
  {S.}~\bibnamefont {Herbers}}, \bibinfo {author} {\bibfnamefont
  {S.}~\bibnamefont {H\"afner}}, \bibinfo {author} {\bibfnamefont
  {U.}~\bibnamefont {Sterr}},\ and\ \bibinfo {author} {\bibfnamefont
  {C.}~\bibnamefont {Lisdat}},\ }\bibfield  {title} {\bibinfo {title} {Towards
  an optical clock for space: {Compact}, high-performance optical lattice clock
  based on bosonic atoms},\ }\href {https://doi.org/10.1103/PhysRevA.98.053443}
  {\bibfield  {journal} {\bibinfo  {journal} {Phys. Rev. A}\ }\textbf {\bibinfo
  {volume} {98}},\ \bibinfo {pages} {053443} (\bibinfo {year}
  {2018})}\BibitemShut {NoStop}%
\bibitem [{\citenamefont {Gellesch}\ \emph {et~al.}(2020)\citenamefont
  {Gellesch}, \citenamefont {Jones}, \citenamefont {Barron}, \citenamefont
  {Singh}, \citenamefont {Sun}, \citenamefont {Bongs},\ and\ \citenamefont
  {Singh}}]{gel20}%
  \BibitemOpen
  \bibfield  {author} {\bibinfo {author} {\bibfnamefont {M.}~\bibnamefont
  {Gellesch}}, \bibinfo {author} {\bibfnamefont {J.}~\bibnamefont {Jones}},
  \bibinfo {author} {\bibfnamefont {R.}~\bibnamefont {Barron}}, \bibinfo
  {author} {\bibfnamefont {A.}~\bibnamefont {Singh}}, \bibinfo {author}
  {\bibfnamefont {Q.}~\bibnamefont {Sun}}, \bibinfo {author} {\bibfnamefont
  {K.}~\bibnamefont {Bongs}},\ and\ \bibinfo {author} {\bibfnamefont
  {Y.}~\bibnamefont {Singh}},\ }\bibfield  {title} {\bibinfo {title}
  {Transportable optical atomic clocks for use in out-of-the-lab
  environments},\ }\href {https://doi.org/10.1515/aot-2020-0023} {\bibfield
  {journal} {\bibinfo  {journal} {Adv. Opt. Technol.}\ }\textbf {\bibinfo
  {volume} {9}},\ \bibinfo {pages} {313} (\bibinfo {year} {2020})}\BibitemShut
  {NoStop}%
\bibitem [{\citenamefont {Ohmae}\ \emph {et~al.}(2021)\citenamefont {Ohmae},
  \citenamefont {Takamoto}, \citenamefont {Takahashi}, \citenamefont {Kokubun},
  \citenamefont {Araki}, \citenamefont {Hinton}, \citenamefont {Ushijima},
  \citenamefont {Muramatsu}, \citenamefont {Furumiya}, \citenamefont {Sakai},
  \citenamefont {Moriya}, \citenamefont {Kamiya}, \citenamefont {Fujii},
  \citenamefont {Muramatsu}, \citenamefont {Shiimado},\ and\ \citenamefont
  {Katori}}]{ohm21}%
  \BibitemOpen
  \bibfield  {author} {\bibinfo {author} {\bibfnamefont {N.}~\bibnamefont
  {Ohmae}}, \bibinfo {author} {\bibfnamefont {M.}~\bibnamefont {Takamoto}},
  \bibinfo {author} {\bibfnamefont {Y.}~\bibnamefont {Takahashi}}, \bibinfo
  {author} {\bibfnamefont {M.}~\bibnamefont {Kokubun}}, \bibinfo {author}
  {\bibfnamefont {K.}~\bibnamefont {Araki}}, \bibinfo {author} {\bibfnamefont
  {A.}~\bibnamefont {Hinton}}, \bibinfo {author} {\bibfnamefont
  {I.}~\bibnamefont {Ushijima}}, \bibinfo {author} {\bibfnamefont
  {T.}~\bibnamefont {Muramatsu}}, \bibinfo {author} {\bibfnamefont
  {T.}~\bibnamefont {Furumiya}}, \bibinfo {author} {\bibfnamefont
  {Y.}~\bibnamefont {Sakai}}, \bibinfo {author} {\bibfnamefont
  {N.}~\bibnamefont {Moriya}}, \bibinfo {author} {\bibfnamefont
  {N.}~\bibnamefont {Kamiya}}, \bibinfo {author} {\bibfnamefont
  {K.}~\bibnamefont {Fujii}}, \bibinfo {author} {\bibfnamefont
  {R.}~\bibnamefont {Muramatsu}}, \bibinfo {author} {\bibfnamefont
  {T.}~\bibnamefont {Shiimado}},\ and\ \bibinfo {author} {\bibfnamefont
  {H.}~\bibnamefont {Katori}},\ }\bibfield  {title} {\bibinfo {title}
  {Transportable strontium optical lattice clocks operated outside laboratory
  at the level of $10^{-18}$ uncertainty},\ }\href
  {https://doi.org/10.1002/qute.202100015} {\bibfield  {journal} {\bibinfo
  {journal} {Adv. Quantum Technol.}\ }\textbf {\bibinfo {volume} {4}},\
  \bibinfo {pages} {2100015} (\bibinfo {year} {2021})}\BibitemShut {NoStop}%
\bibitem [{\citenamefont {Fasano}(2021)}]{fas21a}%
  \BibitemOpen
  \bibfield  {author} {\bibinfo {author} {\bibfnamefont {R.~J.}\ \bibnamefont
  {Fasano}},\ }\emph {\bibinfo {title} {A Transportable Ytterbium Optical
  Lattice Clock}},\ \href {https://scholar.colorado.edu/downloads/1z40kv20z}
  {Ph.D. thesis},\ \bibinfo  {school} {University of Colorado} (\bibinfo {year}
  {2021})\BibitemShut {NoStop}%
\bibitem [{\citenamefont {Stuhler}\ \emph {et~al.}(2021)\citenamefont
  {Stuhler}, \citenamefont {{Abdel Hafiz}}, \citenamefont {Arar}, \citenamefont
  {Bawamia}, \citenamefont {Bergner}, \citenamefont {Biethahn}, \citenamefont
  {Brakhane}, \citenamefont {Didier}, \citenamefont {Fortagh}, \citenamefont
  {Halder}, \citenamefont {Holzwarth}, \citenamefont {Huntemann}, \citenamefont
  {Johanning}, \citenamefont {J\"ordens}, \citenamefont {Kaenders},
  \citenamefont {Karlewski}, \citenamefont {Kienle}, \citenamefont {Krutzik},
  \citenamefont {Lessing}, \citenamefont {Mehlst\"aubler}, \citenamefont
  {Meschede}, \citenamefont {Peik}, \citenamefont {Peters}, \citenamefont
  {Schmidt}, \citenamefont {Siebeneich}, \citenamefont {Tamm}, \citenamefont
  {Vogt}, \citenamefont {Wicht}, \citenamefont {Wunderlich},\ and\
  \citenamefont {Yu}}]{stu21}%
  \BibitemOpen
  \bibfield  {author} {\bibinfo {author} {\bibfnamefont {J.}~\bibnamefont
  {Stuhler}}, \bibinfo {author} {\bibfnamefont {M.}~\bibnamefont {{Abdel
  Hafiz}}}, \bibinfo {author} {\bibfnamefont {B.}~\bibnamefont {Arar}},
  \bibinfo {author} {\bibfnamefont {A.}~\bibnamefont {Bawamia}}, \bibinfo
  {author} {\bibfnamefont {K.}~\bibnamefont {Bergner}}, \bibinfo {author}
  {\bibfnamefont {M.}~\bibnamefont {Biethahn}}, \bibinfo {author}
  {\bibfnamefont {S.}~\bibnamefont {Brakhane}}, \bibinfo {author}
  {\bibfnamefont {A.}~\bibnamefont {Didier}}, \bibinfo {author} {\bibfnamefont
  {J.}~\bibnamefont {Fortagh}}, \bibinfo {author} {\bibfnamefont
  {M.}~\bibnamefont {Halder}}, \bibinfo {author} {\bibfnamefont
  {R.}~\bibnamefont {Holzwarth}}, \bibinfo {author} {\bibfnamefont
  {N.}~\bibnamefont {Huntemann}}, \bibinfo {author} {\bibfnamefont
  {M.}~\bibnamefont {Johanning}}, \bibinfo {author} {\bibfnamefont
  {R.}~\bibnamefont {J\"ordens}}, \bibinfo {author} {\bibfnamefont
  {W.}~\bibnamefont {Kaenders}}, \bibinfo {author} {\bibfnamefont
  {F.}~\bibnamefont {Karlewski}}, \bibinfo {author} {\bibfnamefont
  {F.}~\bibnamefont {Kienle}}, \bibinfo {author} {\bibfnamefont
  {M.}~\bibnamefont {Krutzik}}, \bibinfo {author} {\bibfnamefont
  {M.}~\bibnamefont {Lessing}}, \bibinfo {author} {\bibfnamefont
  {T.}~\bibnamefont {Mehlst\"aubler}}, \bibinfo {author} {\bibfnamefont
  {D.}~\bibnamefont {Meschede}}, \bibinfo {author} {\bibfnamefont
  {E.}~\bibnamefont {Peik}}, \bibinfo {author} {\bibfnamefont {A.}~\bibnamefont
  {Peters}}, \bibinfo {author} {\bibfnamefont {P.}~\bibnamefont {Schmidt}},
  \bibinfo {author} {\bibfnamefont {H.}~\bibnamefont {Siebeneich}}, \bibinfo
  {author} {\bibfnamefont {C.}~\bibnamefont {Tamm}}, \bibinfo {author}
  {\bibfnamefont {E.}~\bibnamefont {Vogt}}, \bibinfo {author} {\bibfnamefont
  {A.}~\bibnamefont {Wicht}}, \bibinfo {author} {\bibfnamefont
  {C.}~\bibnamefont {Wunderlich}},\ and\ \bibinfo {author} {\bibfnamefont
  {J.}~\bibnamefont {Yu}},\ }\bibfield  {title} {\bibinfo {title} {Opticlock:
  Transportable and easy-to-operate optical single-ion clock},\ }\href
  {https://doi.org/10.1016/j.measen.2021.100264} {\bibfield  {journal}
  {\bibinfo  {journal} {Measurement: Sensors}\ }\textbf {\bibinfo {volume}
  {18}},\ \bibinfo {pages} {100264} (\bibinfo {year} {2021})}\BibitemShut
  {NoStop}%
\bibitem [{\citenamefont {Guo}\ \emph {et~al.}(2021)\citenamefont {Guo},
  \citenamefont {Tan}, \citenamefont {Zhou}, \citenamefont {Xia}, \citenamefont
  {Chen}, \citenamefont {Liang}, \citenamefont {Liu}, \citenamefont {Liu},
  \citenamefont {He}, \citenamefont {Zhou}, \citenamefont {Wang}, \citenamefont
  {Shen}, \citenamefont {Zou},\ and\ \citenamefont {Chang}}]{guo21}%
  \BibitemOpen
  \bibfield  {author} {\bibinfo {author} {\bibfnamefont {F.}~\bibnamefont
  {Guo}}, \bibinfo {author} {\bibfnamefont {W.}~\bibnamefont {Tan}}, \bibinfo
  {author} {\bibfnamefont {C.-h.}\ \bibnamefont {Zhou}}, \bibinfo {author}
  {\bibfnamefont {J.}~\bibnamefont {Xia}}, \bibinfo {author} {\bibfnamefont
  {Y.-x.}\ \bibnamefont {Chen}}, \bibinfo {author} {\bibfnamefont
  {T.}~\bibnamefont {Liang}}, \bibinfo {author} {\bibfnamefont
  {Q.}~\bibnamefont {Liu}}, \bibinfo {author} {\bibfnamefont {Y.}~\bibnamefont
  {Liu}}, \bibinfo {author} {\bibfnamefont {D.-j.}\ \bibnamefont {He}},
  \bibinfo {author} {\bibfnamefont {Y.-z.}\ \bibnamefont {Zhou}}, \bibinfo
  {author} {\bibfnamefont {W.-h.}\ \bibnamefont {Wang}}, \bibinfo {author}
  {\bibfnamefont {Y.}~\bibnamefont {Shen}}, \bibinfo {author} {\bibfnamefont
  {H.-x.}\ \bibnamefont {Zou}},\ and\ \bibinfo {author} {\bibfnamefont
  {H.}~\bibnamefont {Chang}},\ }\bibfield  {title} {\bibinfo {title} {A
  proof-of-concept model of compact and high-performance {$^{87}$Sr} optical
  lattice clock for space},\ }\href {https://doi.org/10.1063/5.0064087}
  {\bibfield  {journal} {\bibinfo  {journal} {AIP Advances}\ }\textbf {\bibinfo
  {volume} {11}},\ \bibinfo {pages} {125116} (\bibinfo {year}
  {2021})}\BibitemShut {NoStop}%
\bibitem [{\citenamefont {Zeng}\ \emph {et~al.}(2023)\citenamefont {Zeng},
  \citenamefont {Huang}, \citenamefont {Zhang}, \citenamefont {Hao},
  \citenamefont {Ma}, \citenamefont {Hu}, \citenamefont {Zhang}, \citenamefont
  {Chen}, \citenamefont {Wang}, \citenamefont {Guan},\ and\ \citenamefont
  {Gao}}]{zen23}%
  \BibitemOpen
  \bibfield  {author} {\bibinfo {author} {\bibfnamefont {M.}~\bibnamefont
  {Zeng}}, \bibinfo {author} {\bibfnamefont {Y.}~\bibnamefont {Huang}},
  \bibinfo {author} {\bibfnamefont {B.}~\bibnamefont {Zhang}}, \bibinfo
  {author} {\bibfnamefont {Y.}~\bibnamefont {Hao}}, \bibinfo {author}
  {\bibfnamefont {Z.}~\bibnamefont {Ma}}, \bibinfo {author} {\bibfnamefont
  {R.}~\bibnamefont {Hu}}, \bibinfo {author} {\bibfnamefont {H.}~\bibnamefont
  {Zhang}}, \bibinfo {author} {\bibfnamefont {Z.}~\bibnamefont {Chen}},
  \bibinfo {author} {\bibfnamefont {M.}~\bibnamefont {Wang}}, \bibinfo {author}
  {\bibfnamefont {H.}~\bibnamefont {Guan}},\ and\ \bibinfo {author}
  {\bibfnamefont {K.}~\bibnamefont {Gao}},\ }\bibfield  {title} {\bibinfo
  {title} {Toward a transportable {Ca}$^+$ optical clock with a systematic
  uncertainty of $4.8 \times 10^{-18}$},\ }\href
  {https://doi.org/10.1103/PhysRevApplied.19.064004} {\bibfield  {journal}
  {\bibinfo  {journal} {Phys. Rev. Appl.}\ }\textbf {\bibinfo {volume} {19}},\
  \bibinfo {pages} {064004} (\bibinfo {year} {2023})}\BibitemShut {NoStop}%
\bibitem [{\citenamefont {Falke}\ \emph {et~al.}(2014)\citenamefont {Falke},
  \citenamefont {Lemke}, \citenamefont {Grebing}, \citenamefont {Lipphardt},
  \citenamefont {Weyers}, \citenamefont {Gerginov}, \citenamefont {Huntemann},
  \citenamefont {Hagemann}, \citenamefont {Al-Masoudi}, \citenamefont
  {H{\"a}fner}, \citenamefont {Vogt}, \citenamefont {Sterr},\ and\
  \citenamefont {Lisdat}}]{fal14}%
  \BibitemOpen
  \bibfield  {author} {\bibinfo {author} {\bibfnamefont {S.}~\bibnamefont
  {Falke}}, \bibinfo {author} {\bibfnamefont {N.}~\bibnamefont {Lemke}},
  \bibinfo {author} {\bibfnamefont {C.}~\bibnamefont {Grebing}}, \bibinfo
  {author} {\bibfnamefont {B.}~\bibnamefont {Lipphardt}}, \bibinfo {author}
  {\bibfnamefont {S.}~\bibnamefont {Weyers}}, \bibinfo {author} {\bibfnamefont
  {V.}~\bibnamefont {Gerginov}}, \bibinfo {author} {\bibfnamefont
  {N.}~\bibnamefont {Huntemann}}, \bibinfo {author} {\bibfnamefont
  {C.}~\bibnamefont {Hagemann}}, \bibinfo {author} {\bibfnamefont
  {A.}~\bibnamefont {Al-Masoudi}}, \bibinfo {author} {\bibfnamefont
  {S.}~\bibnamefont {H{\"a}fner}}, \bibinfo {author} {\bibfnamefont
  {S.}~\bibnamefont {Vogt}}, \bibinfo {author} {\bibfnamefont {U.}~\bibnamefont
  {Sterr}},\ and\ \bibinfo {author} {\bibfnamefont {C.}~\bibnamefont
  {Lisdat}},\ }\bibfield  {title} {\bibinfo {title} {A strontium lattice clock
  with $3 \times 10^{-17}$ inaccuracy and its frequency},\ }\href
  {https://doi.org/10.1088/1367-2630/16/7/073023} {\bibfield  {journal}
  {\bibinfo  {journal} {New J. Phys.}\ }\textbf {\bibinfo {volume} {16}},\
  \bibinfo {pages} {073023} (\bibinfo {year} {2014})}\BibitemShut {NoStop}%
\bibitem [{\citenamefont {Schwarz}\ \emph {et~al.}(2020)\citenamefont
  {Schwarz}, \citenamefont {D\"{o}rscher}, \citenamefont {Al-Masoudi},
  \citenamefont {Benkler}, \citenamefont {Legero}, \citenamefont {Sterr},
  \citenamefont {Weyers}, \citenamefont {Rahm}, \citenamefont {Lipphardt},\
  and\ \citenamefont {Lisdat}}]{sch20d}%
  \BibitemOpen
  \bibfield  {author} {\bibinfo {author} {\bibfnamefont {R.}~\bibnamefont
  {Schwarz}}, \bibinfo {author} {\bibfnamefont {S.}~\bibnamefont
  {D\"{o}rscher}}, \bibinfo {author} {\bibfnamefont {A.}~\bibnamefont
  {Al-Masoudi}}, \bibinfo {author} {\bibfnamefont {E.}~\bibnamefont {Benkler}},
  \bibinfo {author} {\bibfnamefont {T.}~\bibnamefont {Legero}}, \bibinfo
  {author} {\bibfnamefont {U.}~\bibnamefont {Sterr}}, \bibinfo {author}
  {\bibfnamefont {S.}~\bibnamefont {Weyers}}, \bibinfo {author} {\bibfnamefont
  {J.}~\bibnamefont {Rahm}}, \bibinfo {author} {\bibfnamefont {B.}~\bibnamefont
  {Lipphardt}},\ and\ \bibinfo {author} {\bibfnamefont {C.}~\bibnamefont
  {Lisdat}},\ }\bibfield  {title} {\bibinfo {title} {Long term measurement of
  the {$^{87}$Sr} clock frequency at the limit of primary {Cs} clocks},\ }\href
  {https://doi.org/10.1103/PhysRevResearch.2.033242} {\bibfield  {journal}
  {\bibinfo  {journal} {Phys. Rev. Res.}\ }\textbf {\bibinfo {volume} {2}},\
  \bibinfo {pages} {033242} (\bibinfo {year} {2020})}\BibitemShut {NoStop}%
\bibitem [{\citenamefont {Raupach}\ \emph {et~al.}(2015)\citenamefont
  {Raupach}, \citenamefont {Koczwara},\ and\ \citenamefont {Grosche}}]{rau15}%
  \BibitemOpen
  \bibfield  {author} {\bibinfo {author} {\bibfnamefont {S.~M.~F.}\
  \bibnamefont {Raupach}}, \bibinfo {author} {\bibfnamefont {A.}~\bibnamefont
  {Koczwara}},\ and\ \bibinfo {author} {\bibfnamefont {G.}~\bibnamefont
  {Grosche}},\ }\bibfield  {title} {\bibinfo {title} {Brillouin amplification
  supports $1\times10^{-20}$ accuracy in optical frequency transfer over
  1400~km of underground fibre},\ }\href
  {https://doi.org/10.1103/PhysRevA.92.021801} {\bibfield  {journal} {\bibinfo
  {journal} {Phys. Rev. A}\ }\textbf {\bibinfo {volume} {92}},\ \bibinfo
  {pages} {021801(R)} (\bibinfo {year} {2015})}\BibitemShut {NoStop}%
\bibitem [{\citenamefont {Kramer}\ and\ \citenamefont
  {Klische}(2004)}]{kra04a}%
  \BibitemOpen
  \bibfield  {author} {\bibinfo {author} {\bibfnamefont {G.}~\bibnamefont
  {Kramer}}\ and\ \bibinfo {author} {\bibfnamefont {W.}~\bibnamefont
  {Klische}},\ }\bibfield  {title} {\bibinfo {title} {Extra high precision
  digital phase recorder},\ }in\ \href {https://doi.org/10.1049/cp:20040866}
  {\emph {\bibinfo {booktitle} {Proceedings of the 18th European Frequency and
  Time Forum, Guildford, UK}}}\ (\bibinfo  {publisher} {IET},\ \bibinfo
  {address} {London, UK},\ \bibinfo {year} {2004})\ pp.\ \bibinfo {pages}
  {595--602}\BibitemShut {NoStop}%
\bibitem [{\citenamefont {Benkler}\ \emph {et~al.}(2015)\citenamefont
  {Benkler}, \citenamefont {Lisdat},\ and\ \citenamefont {Sterr}}]{ben15}%
  \BibitemOpen
  \bibfield  {author} {\bibinfo {author} {\bibfnamefont {E.}~\bibnamefont
  {Benkler}}, \bibinfo {author} {\bibfnamefont {C.}~\bibnamefont {Lisdat}},\
  and\ \bibinfo {author} {\bibfnamefont {U.}~\bibnamefont {Sterr}},\ }\bibfield
   {title} {\bibinfo {title} {On the relation between uncertainties of weighted
  frequency averages and the various types of {Allan} deviations},\ }\href
  {https://doi.org/10.1088/0026-1394/52/4/565} {\bibfield  {journal} {\bibinfo
  {journal} {Metrologia}\ }\textbf {\bibinfo {volume} {52}},\ \bibinfo {pages}
  {565} (\bibinfo {year} {2015})}\BibitemShut {NoStop}%
\bibitem [{\citenamefont {Ushijima}\ \emph {et~al.}(2018)\citenamefont
  {Ushijima}, \citenamefont {Takamoto},\ and\ \citenamefont {Katori}}]{ush18}%
  \BibitemOpen
  \bibfield  {author} {\bibinfo {author} {\bibfnamefont {I.}~\bibnamefont
  {Ushijima}}, \bibinfo {author} {\bibfnamefont {M.}~\bibnamefont {Takamoto}},\
  and\ \bibinfo {author} {\bibfnamefont {H.}~\bibnamefont {Katori}},\
  }\bibfield  {title} {\bibinfo {title} {Operational magic intensity for {Sr}
  optical lattice clocks},\ }\href
  {https://doi.org/10.1103/PhysRevLett.121.263202} {\bibfield  {journal}
  {\bibinfo  {journal} {Phys. Rev. Lett.}\ }\textbf {\bibinfo {volume} {121}},\
  \bibinfo {pages} {263202} (\bibinfo {year} {2018})}\BibitemShut {NoStop}%
\bibitem [{Sup()}]{SupplementalMaterial}%
  \BibitemOpen
  \bibinfo {howpublished} {See Supplemental Material at \url{http://link.aps.org/supplemental/10.1103/PhysRevApplied.21.L061001} (and below) for additional information on the
  consideration of correlations in weighted averaging and the evaluation of
  systematic frequency shifts in the transportable clock.}\BibitemShut {Stop}%
\bibitem [{GUM()}]{gum08}%
  \BibitemOpen
  GUM,\ \href
  {http://www.bipm.org/utils/common/documents/jcgm/JCGM_100_2008_E.pdf}
  {\bibinfo {title} {Evaluation of measurement data -- guide to the expression
  of uncertainty in measurement}},\ \bibinfo {howpublished} {JCGM 100:2008}
  (\bibinfo {year} {2008})\BibitemShut {NoStop}%
\bibitem [{\citenamefont {Middelmann}\ \emph {et~al.}(2012)\citenamefont
  {Middelmann}, \citenamefont {Falke}, \citenamefont {Lisdat},\ and\
  \citenamefont {Sterr}}]{mid12a}%
  \BibitemOpen
  \bibfield  {author} {\bibinfo {author} {\bibfnamefont {T.}~\bibnamefont
  {Middelmann}}, \bibinfo {author} {\bibfnamefont {S.}~\bibnamefont {Falke}},
  \bibinfo {author} {\bibfnamefont {C.}~\bibnamefont {Lisdat}},\ and\ \bibinfo
  {author} {\bibfnamefont {U.}~\bibnamefont {Sterr}},\ }\bibfield  {title}
  {\bibinfo {title} {High accuracy correction of blackbody radiation shift in
  an optical lattice clock},\ }\href
  {https://doi.org/10.1103/PhysRevLett.109.263004} {\bibfield  {journal}
  {\bibinfo  {journal} {Phys. Rev. Lett.}\ }\textbf {\bibinfo {volume} {109}},\
  \bibinfo {pages} {263004} (\bibinfo {year} {2012})}\BibitemShut {NoStop}%
\bibitem [{\citenamefont {Nicholson}\ \emph {et~al.}(2015)\citenamefont
  {Nicholson}, \citenamefont {Campbell}, \citenamefont {Hutson}, \citenamefont
  {Marti}, \citenamefont {Bloom}, \citenamefont {McNally}, \citenamefont
  {Zhang}, \citenamefont {Barrett}, \citenamefont {Safronova}, \citenamefont
  {Strouse}, \citenamefont {Tew},\ and\ \citenamefont {Ye}}]{nic15}%
  \BibitemOpen
  \bibfield  {author} {\bibinfo {author} {\bibfnamefont {T.~L.}\ \bibnamefont
  {Nicholson}}, \bibinfo {author} {\bibfnamefont {S.~L.}\ \bibnamefont
  {Campbell}}, \bibinfo {author} {\bibfnamefont {R.~B.}\ \bibnamefont
  {Hutson}}, \bibinfo {author} {\bibfnamefont {G.~E.}\ \bibnamefont {Marti}},
  \bibinfo {author} {\bibfnamefont {B.~J.}\ \bibnamefont {Bloom}}, \bibinfo
  {author} {\bibfnamefont {R.~L.}\ \bibnamefont {McNally}}, \bibinfo {author}
  {\bibfnamefont {W.}~\bibnamefont {Zhang}}, \bibinfo {author} {\bibfnamefont
  {M.~D.}\ \bibnamefont {Barrett}}, \bibinfo {author} {\bibfnamefont {M.~S.}\
  \bibnamefont {Safronova}}, \bibinfo {author} {\bibfnamefont {G.~F.}\
  \bibnamefont {Strouse}}, \bibinfo {author} {\bibfnamefont {W.~L.}\
  \bibnamefont {Tew}},\ and\ \bibinfo {author} {\bibfnamefont {J.}~\bibnamefont
  {Ye}},\ }\bibfield  {title} {\bibinfo {title} {Systematic evaluation of an
  atomic clock at $2 \times 10^{-18}$ total uncertainty},\ }\href
  {https://doi.org/10.1038/ncomms7896} {\bibfield  {journal} {\bibinfo
  {journal} {Nature Commun.}\ }\textbf {\bibinfo {volume} {6}},\ \bibinfo
  {pages} {6896} (\bibinfo {year} {2015})}\BibitemShut {NoStop}%
\bibitem [{\citenamefont {Lisdat}\ \emph {et~al.}(2021)\citenamefont {Lisdat},
  \citenamefont {D\"orscher}, \citenamefont {Nosske},\ and\ \citenamefont
  {Sterr}}]{lis21a}%
  \BibitemOpen
  \bibfield  {author} {\bibinfo {author} {\bibfnamefont {C.}~\bibnamefont
  {Lisdat}}, \bibinfo {author} {\bibfnamefont {S.}~\bibnamefont {D\"orscher}},
  \bibinfo {author} {\bibfnamefont {I.}~\bibnamefont {Nosske}},\ and\ \bibinfo
  {author} {\bibfnamefont {U.}~\bibnamefont {Sterr}},\ }\bibfield  {title}
  {\bibinfo {title} {Blackbody radiation shift in strontium lattice clocks
  revisited},\ }\href {https://doi.org/10.1103/PhysRevResearch.3.L042036}
  {\bibfield  {journal} {\bibinfo  {journal} {Phys. Rev. Res.}\ }\textbf
  {\bibinfo {volume} {3}},\ \bibinfo {pages} {L042036} (\bibinfo {year}
  {2021})}\BibitemShut {NoStop}%
\bibitem [{\citenamefont {Al-Masoudi}\ \emph {et~al.}(2015)\citenamefont
  {Al-Masoudi}, \citenamefont {D\"orscher}, \citenamefont {H\"afner},
  \citenamefont {Sterr},\ and\ \citenamefont {Lisdat}}]{alm15}%
  \BibitemOpen
  \bibfield  {author} {\bibinfo {author} {\bibfnamefont {A.}~\bibnamefont
  {Al-Masoudi}}, \bibinfo {author} {\bibfnamefont {S.}~\bibnamefont
  {D\"orscher}}, \bibinfo {author} {\bibfnamefont {S.}~\bibnamefont
  {H\"afner}}, \bibinfo {author} {\bibfnamefont {U.}~\bibnamefont {Sterr}},\
  and\ \bibinfo {author} {\bibfnamefont {C.}~\bibnamefont {Lisdat}},\
  }\bibfield  {title} {\bibinfo {title} {Noise and instability of an optical
  lattice clock},\ }\href {https://doi.org/10.1103/PhysRevA.92.063814}
  {\bibfield  {journal} {\bibinfo  {journal} {Phys. Rev. A}\ }\textbf {\bibinfo
  {volume} {92}},\ \bibinfo {pages} {063814} (\bibinfo {year}
  {2015})}\BibitemShut {NoStop}%
\bibitem [{\citenamefont {Boyd}\ \emph {et~al.}(2007)\citenamefont {Boyd},
  \citenamefont {Zelevinsky}, \citenamefont {Ludlow}, \citenamefont {Blatt},
  \citenamefont {Zanon-Willette}, \citenamefont {Foreman},\ and\ \citenamefont
  {Ye}}]{boy07a}%
  \BibitemOpen
  \bibfield  {author} {\bibinfo {author} {\bibfnamefont {M.~M.}\ \bibnamefont
  {Boyd}}, \bibinfo {author} {\bibfnamefont {T.}~\bibnamefont {Zelevinsky}},
  \bibinfo {author} {\bibfnamefont {A.~D.}\ \bibnamefont {Ludlow}}, \bibinfo
  {author} {\bibfnamefont {S.}~\bibnamefont {Blatt}}, \bibinfo {author}
  {\bibfnamefont {T.}~\bibnamefont {Zanon-Willette}}, \bibinfo {author}
  {\bibfnamefont {S.~M.}\ \bibnamefont {Foreman}},\ and\ \bibinfo {author}
  {\bibfnamefont {J.}~\bibnamefont {Ye}},\ }\bibfield  {title} {\bibinfo
  {title} {Nuclear spin effects in optical lattice clocks},\ }\href
  {https://doi.org/10.1103/PhysRevA.76.022510} {\bibfield  {journal} {\bibinfo
  {journal} {Phys. Rev. A}\ }\textbf {\bibinfo {volume} {76}},\ \bibinfo
  {pages} {022510} (\bibinfo {year} {2007})}\BibitemShut {NoStop}%
\bibitem [{\citenamefont {D{\"o}rscher}\ \emph {et~al.}(2018)\citenamefont
  {D{\"o}rscher}, \citenamefont {Schwarz}, \citenamefont {Al-Masoudi},
  \citenamefont {Falke}, \citenamefont {Sterr},\ and\ \citenamefont
  {Lisdat}}]{doe18}%
  \BibitemOpen
  \bibfield  {author} {\bibinfo {author} {\bibfnamefont {S.}~\bibnamefont
  {D{\"o}rscher}}, \bibinfo {author} {\bibfnamefont {R.}~\bibnamefont
  {Schwarz}}, \bibinfo {author} {\bibfnamefont {A.}~\bibnamefont {Al-Masoudi}},
  \bibinfo {author} {\bibfnamefont {S.}~\bibnamefont {Falke}}, \bibinfo
  {author} {\bibfnamefont {U.}~\bibnamefont {Sterr}},\ and\ \bibinfo {author}
  {\bibfnamefont {C.}~\bibnamefont {Lisdat}},\ }\bibfield  {title} {\bibinfo
  {title} {Lattice-induced photon scattering in an optical lattice clock},\
  }\href {https://doi.org/10.1103/PhysRevA.97.063419} {\bibfield  {journal}
  {\bibinfo  {journal} {Phys. Rev. A}\ }\textbf {\bibinfo {volume} {97}},\
  \bibinfo {pages} {063419} (\bibinfo {year} {2018})}\BibitemShut {NoStop}%
\bibitem [{\citenamefont {Muniz}\ \emph {et~al.}(2021)\citenamefont {Muniz},
  \citenamefont {Young}, \citenamefont {Cline},\ and\ \citenamefont
  {Thompson}}]{mun21}%
  \BibitemOpen
  \bibfield  {author} {\bibinfo {author} {\bibfnamefont {J.~A.}\ \bibnamefont
  {Muniz}}, \bibinfo {author} {\bibfnamefont {D.~J.}\ \bibnamefont {Young}},
  \bibinfo {author} {\bibfnamefont {J.~R.~K.}\ \bibnamefont {Cline}},\ and\
  \bibinfo {author} {\bibfnamefont {J.~K.}\ \bibnamefont {Thompson}},\
  }\bibfield  {title} {\bibinfo {title} {Cavity-{QED} measurements of the
  $^{87}\mathrm{Sr}$ millihertz optical clock transition and determination of
  its natural linewidth},\ }\href
  {https://doi.org/10.1103/PhysRevResearch.3.023152} {\bibfield  {journal}
  {\bibinfo  {journal} {Phys. Rev. Res.}\ }\textbf {\bibinfo {volume} {3}},\
  \bibinfo {pages} {023152} (\bibinfo {year} {2021})}\BibitemShut {NoStop}%
\bibitem [{\citenamefont {Baillard}\ \emph {et~al.}(2007)\citenamefont
  {Baillard}, \citenamefont {Fouch\'e}, \citenamefont {Le~Targat},
  \citenamefont {Westergaard}, \citenamefont {Lecallier}, \citenamefont
  {Le~Coq}, \citenamefont {Rovera}, \citenamefont {Bize},\ and\ \citenamefont
  {Lemonde}}]{bai07}%
  \BibitemOpen
  \bibfield  {author} {\bibinfo {author} {\bibfnamefont {X.}~\bibnamefont
  {Baillard}}, \bibinfo {author} {\bibfnamefont {M.}~\bibnamefont {Fouch\'e}},
  \bibinfo {author} {\bibfnamefont {R.}~\bibnamefont {Le~Targat}}, \bibinfo
  {author} {\bibfnamefont {P.~G.}\ \bibnamefont {Westergaard}}, \bibinfo
  {author} {\bibfnamefont {A.}~\bibnamefont {Lecallier}}, \bibinfo {author}
  {\bibfnamefont {Y.}~\bibnamefont {Le~Coq}}, \bibinfo {author} {\bibfnamefont
  {G.~D.}\ \bibnamefont {Rovera}}, \bibinfo {author} {\bibfnamefont
  {S.}~\bibnamefont {Bize}},\ and\ \bibinfo {author} {\bibfnamefont
  {P.}~\bibnamefont {Lemonde}},\ }\bibfield  {title} {\bibinfo {title}
  {Accuracy evaluation of an optical lattice clock with bosonic atoms},\ }\href
  {http://www.opticsinfobase.org/abstract.cfm?URI=ol-32-13-1812} {\bibfield
  {journal} {\bibinfo  {journal} {Opt. Lett.}\ }\textbf {\bibinfo {volume}
  {32}},\ \bibinfo {pages} {1812} (\bibinfo {year} {2007})}\BibitemShut
  {NoStop}%
\bibitem [{\citenamefont {Shi}\ \emph {et~al.}(2015)\citenamefont {Shi},
  \citenamefont {Robyr}, \citenamefont {Eismann}, \citenamefont {Zawada},
  \citenamefont {Lorini}, \citenamefont {Le~Targat},\ and\ \citenamefont
  {Lodewyck}}]{shi15}%
  \BibitemOpen
  \bibfield  {author} {\bibinfo {author} {\bibfnamefont {C.}~\bibnamefont
  {Shi}}, \bibinfo {author} {\bibfnamefont {J.-L.}\ \bibnamefont {Robyr}},
  \bibinfo {author} {\bibfnamefont {U.}~\bibnamefont {Eismann}}, \bibinfo
  {author} {\bibfnamefont {M.}~\bibnamefont {Zawada}}, \bibinfo {author}
  {\bibfnamefont {L.}~\bibnamefont {Lorini}}, \bibinfo {author} {\bibfnamefont
  {R.}~\bibnamefont {Le~Targat}},\ and\ \bibinfo {author} {\bibfnamefont
  {J.}~\bibnamefont {Lodewyck}},\ }\bibfield  {title} {\bibinfo {title}
  {Polarizabilities of the {$^{87}$Sr} clock transition},\ }\href
  {https://doi.org/10.1103/PhysRevA.92.012516} {\bibfield  {journal} {\bibinfo
  {journal} {Phys. Rev. A}\ }\textbf {\bibinfo {volume} {92}},\ \bibinfo
  {pages} {012516} (\bibinfo {year} {2015})}\BibitemShut {NoStop}%
\bibitem [{\citenamefont {Bothwell}\ \emph {et~al.}(2019)\citenamefont
  {Bothwell}, \citenamefont {Kedar}, \citenamefont {Oelker}, \citenamefont
  {Robinson}, \citenamefont {Bromley}, \citenamefont {Tew}, \citenamefont
  {Ye},\ and\ \citenamefont {Kennedy}}]{bot19}%
  \BibitemOpen
  \bibfield  {author} {\bibinfo {author} {\bibfnamefont {T.}~\bibnamefont
  {Bothwell}}, \bibinfo {author} {\bibfnamefont {D.}~\bibnamefont {Kedar}},
  \bibinfo {author} {\bibfnamefont {E.}~\bibnamefont {Oelker}}, \bibinfo
  {author} {\bibfnamefont {J.~M.}\ \bibnamefont {Robinson}}, \bibinfo {author}
  {\bibfnamefont {S.~L.}\ \bibnamefont {Bromley}}, \bibinfo {author}
  {\bibfnamefont {W.~L.}\ \bibnamefont {Tew}}, \bibinfo {author} {\bibfnamefont
  {J.}~\bibnamefont {Ye}},\ and\ \bibinfo {author} {\bibfnamefont {C.~J.}\
  \bibnamefont {Kennedy}},\ }\bibfield  {title} {\bibinfo {title} {{JILA} {SrI}
  optical lattice clock with uncertainty of $2.0 \times 10^{-18}$},\ }\href
  {https://doi.org/10.1088/1681-7575/ab4089} {\bibfield  {journal} {\bibinfo
  {journal} {Metrologia}\ }\textbf {\bibinfo {volume} {56}},\ \bibinfo {pages}
  {065004} (\bibinfo {year} {2019})}\BibitemShut {NoStop}%
\bibitem [{\citenamefont {Alves}\ \emph {et~al.}(2019)\citenamefont {Alves},
  \citenamefont {Foucault}, \citenamefont {Vallet},\ and\ \citenamefont
  {Lodewyck}}]{alv19}%
  \BibitemOpen
  \bibfield  {author} {\bibinfo {author} {\bibfnamefont {B.~X.~R.}\
  \bibnamefont {Alves}}, \bibinfo {author} {\bibfnamefont {Y.}~\bibnamefont
  {Foucault}}, \bibinfo {author} {\bibfnamefont {G.}~\bibnamefont {Vallet}},\
  and\ \bibinfo {author} {\bibfnamefont {J.}~\bibnamefont {Lodewyck}},\
  }\bibfield  {title} {\bibinfo {title} {Background gas collision frequency
  shift on lattice-trapped strontium atoms},\ }in\ \href
  {https://doi.org/10.1109/FCS.2019.8856042} {\emph {\bibinfo {booktitle} {2019
  Joint Conference of the IEEE International Frequency Control Symposium and
  European Frequency and Time Forum (EFTF/IFC)}}}\ (\bibinfo {organization}
  {IEEE},\ \bibinfo {address} {Orlando, FL, USA},\ \bibinfo {year} {2019})\
  pp.\ \bibinfo {pages} {1--2}\BibitemShut {NoStop}%
\bibitem [{\citenamefont {Falke}\ \emph {et~al.}(2011)\citenamefont {Falke},
  \citenamefont {Schnatz}, \citenamefont {Vellore~Winfred}, \citenamefont
  {Middelmann}, \citenamefont {Vogt}, \citenamefont {Weyers}, \citenamefont
  {Lipphardt}, \citenamefont {Grosche}, \citenamefont {Riehle}, \citenamefont
  {Sterr},\ and\ \citenamefont {Lisdat}}]{fal11}%
  \BibitemOpen
  \bibfield  {author} {\bibinfo {author} {\bibfnamefont {S.}~\bibnamefont
  {Falke}}, \bibinfo {author} {\bibfnamefont {H.}~\bibnamefont {Schnatz}},
  \bibinfo {author} {\bibfnamefont {J.~S.~R.}\ \bibnamefont {Vellore~Winfred}},
  \bibinfo {author} {\bibfnamefont {T.}~\bibnamefont {Middelmann}}, \bibinfo
  {author} {\bibfnamefont {S.}~\bibnamefont {Vogt}}, \bibinfo {author}
  {\bibfnamefont {S.}~\bibnamefont {Weyers}}, \bibinfo {author} {\bibfnamefont
  {B.}~\bibnamefont {Lipphardt}}, \bibinfo {author} {\bibfnamefont
  {G.}~\bibnamefont {Grosche}}, \bibinfo {author} {\bibfnamefont
  {F.}~\bibnamefont {Riehle}}, \bibinfo {author} {\bibfnamefont
  {U.}~\bibnamefont {Sterr}},\ and\ \bibinfo {author} {\bibfnamefont
  {C.}~\bibnamefont {Lisdat}},\ }\bibfield  {title} {\bibinfo {title} {The
  $^{87}${S}r optical frequency standard at {PTB}},\ }\href
  {https://doi.org/doi:10.1088/0026-1394/48/5/022} {\bibfield  {journal}
  {\bibinfo  {journal} {Metrologia}\ }\textbf {\bibinfo {volume} {48}},\
  \bibinfo {pages} {399} (\bibinfo {year} {2011})}\BibitemShut {NoStop}%
\bibitem [{\citenamefont {Lemonde}\ and\ \citenamefont {Wolf}(2005)}]{lem05}%
  \BibitemOpen
  \bibfield  {author} {\bibinfo {author} {\bibfnamefont {P.}~\bibnamefont
  {Lemonde}}\ and\ \bibinfo {author} {\bibfnamefont {P.}~\bibnamefont {Wolf}},\
  }\bibfield  {title} {\bibinfo {title} {Optical lattice clock with atoms
  confined in a shallow trap},\ }\href
  {https://doi.org/10.1103/PhysRevA.72.033409} {\bibfield  {journal} {\bibinfo
  {journal} {Phys. Rev. A}\ }\textbf {\bibinfo {volume} {72}},\ \bibinfo
  {pages} {033409} (\bibinfo {year} {2005})}\BibitemShut {NoStop}%
\bibitem [{\citenamefont {Kacker}\ \emph {et~al.}(2002)\citenamefont {Kacker},
  \citenamefont {Datla},\ and\ \citenamefont {Parr}}]{kac02}%
  \BibitemOpen
  \bibfield  {author} {\bibinfo {author} {\bibfnamefont {R.}~\bibnamefont
  {Kacker}}, \bibinfo {author} {\bibfnamefont {R.}~\bibnamefont {Datla}},\ and\
  \bibinfo {author} {\bibfnamefont {A.}~\bibnamefont {Parr}},\ }\bibfield
  {title} {\bibinfo {title} {Combined result and associated uncertainty from
  interlaboratory evaluations based on the iso guide},\ }\href
  {https://doi.org/10.1088/0026-1394/39/3/5} {\bibfield  {journal} {\bibinfo
  {journal} {Metrologia}\ }\textbf {\bibinfo {volume} {39}},\ \bibinfo {pages}
  {279} (\bibinfo {year} {2002})}\BibitemShut {NoStop}%
\bibitem [{\citenamefont {Benkler}\ \emph {et~al.}(2019)\citenamefont
  {Benkler}, \citenamefont {Lipphardt}, \citenamefont {Puppe}, \citenamefont
  {Wilk}, \citenamefont {Rohde},\ and\ \citenamefont {Sterr}}]{ben19}%
  \BibitemOpen
  \bibfield  {author} {\bibinfo {author} {\bibfnamefont {E.}~\bibnamefont
  {Benkler}}, \bibinfo {author} {\bibfnamefont {B.}~\bibnamefont {Lipphardt}},
  \bibinfo {author} {\bibfnamefont {T.}~\bibnamefont {Puppe}}, \bibinfo
  {author} {\bibfnamefont {R.}~\bibnamefont {Wilk}}, \bibinfo {author}
  {\bibfnamefont {F.}~\bibnamefont {Rohde}},\ and\ \bibinfo {author}
  {\bibfnamefont {U.}~\bibnamefont {Sterr}},\ }\bibfield  {title} {\bibinfo
  {title} {End-to-end topology for fiber comb based optical frequency transfer
  at the $10^{-21}$ level},\ }\href {https://doi.org/10.1364/OE.27.036886}
  {\bibfield  {journal} {\bibinfo  {journal} {Opt. Express}\ }\textbf {\bibinfo
  {volume} {27}},\ \bibinfo {pages} {36886} (\bibinfo {year} {2019})},\
  \bibinfo {note} {also see erratum \cite{ben20}}\BibitemShut {NoStop}%
\bibitem [{\citenamefont {Giunta}\ \emph {et~al.}(2020)\citenamefont {Giunta},
  \citenamefont {H\"ansel}, \citenamefont {Fischer}, \citenamefont {Lezius},
  \citenamefont {Udem},\ and\ \citenamefont {Holzwarth}}]{giu19}%
  \BibitemOpen
  \bibfield  {author} {\bibinfo {author} {\bibfnamefont {M.}~\bibnamefont
  {Giunta}}, \bibinfo {author} {\bibfnamefont {W.}~\bibnamefont {H\"ansel}},
  \bibinfo {author} {\bibfnamefont {M.}~\bibnamefont {Fischer}}, \bibinfo
  {author} {\bibfnamefont {M.}~\bibnamefont {Lezius}}, \bibinfo {author}
  {\bibfnamefont {T.}~\bibnamefont {Udem}},\ and\ \bibinfo {author}
  {\bibfnamefont {R.}~\bibnamefont {Holzwarth}},\ }\bibfield  {title} {\bibinfo
  {title} {Real-time phase tracking for wide-band optical frequency
  measurements at the 20th decimal place},\ }\href
  {https://doi.org/10.1038/s41566-019-0520-5} {\bibfield  {journal} {\bibinfo
  {journal} {Nat. Photonics}\ }\textbf {\bibinfo {volume} {14}},\ \bibinfo
  {pages} {44} (\bibinfo {year} {2020})}\BibitemShut {NoStop}%
\bibitem [{\citenamefont {H{\"a}fner}\ \emph {et~al.}(2020)\citenamefont
  {H{\"a}fner}, \citenamefont {Herbers}, \citenamefont {Vogt}, \citenamefont
  {Lisdat},\ and\ \citenamefont {Sterr}}]{hae20}%
  \BibitemOpen
  \bibfield  {author} {\bibinfo {author} {\bibfnamefont {S.}~\bibnamefont
  {H{\"a}fner}}, \bibinfo {author} {\bibfnamefont {S.}~\bibnamefont {Herbers}},
  \bibinfo {author} {\bibfnamefont {S.}~\bibnamefont {Vogt}}, \bibinfo {author}
  {\bibfnamefont {C.}~\bibnamefont {Lisdat}},\ and\ \bibinfo {author}
  {\bibfnamefont {U.}~\bibnamefont {Sterr}},\ }\bibfield  {title} {\bibinfo
  {title} {Transportable interrogation laser system with an instability of
  $\mathrm{mod}~\sigma_y = 3\times 10^{-16}$},\ }\href
  {https://doi.org/10.1364/OE.390105} {\bibfield  {journal} {\bibinfo
  {journal} {Opt. Express}\ }\textbf {\bibinfo {volume} {28}},\ \bibinfo
  {pages} {16407} (\bibinfo {year} {2020})}\BibitemShut {NoStop}%
\bibitem [{\citenamefont {H{\"a}fner}\ \emph {et~al.}(2015)\citenamefont
  {H{\"a}fner}, \citenamefont {Falke}, \citenamefont {Grebing}, \citenamefont
  {Vogt}, \citenamefont {Legero}, \citenamefont {Merimaa}, \citenamefont
  {Lisdat},\ and\ \citenamefont {Sterr}}]{hae15a}%
  \BibitemOpen
  \bibfield  {author} {\bibinfo {author} {\bibfnamefont {S.}~\bibnamefont
  {H{\"a}fner}}, \bibinfo {author} {\bibfnamefont {S.}~\bibnamefont {Falke}},
  \bibinfo {author} {\bibfnamefont {C.}~\bibnamefont {Grebing}}, \bibinfo
  {author} {\bibfnamefont {S.}~\bibnamefont {Vogt}}, \bibinfo {author}
  {\bibfnamefont {T.}~\bibnamefont {Legero}}, \bibinfo {author} {\bibfnamefont
  {M.}~\bibnamefont {Merimaa}}, \bibinfo {author} {\bibfnamefont
  {C.}~\bibnamefont {Lisdat}},\ and\ \bibinfo {author} {\bibfnamefont
  {U.}~\bibnamefont {Sterr}},\ }\bibfield  {title} {\bibinfo {title} {$8 \times
  10^{-17}$ fractional laser frequency instability with a long room-temperature
  cavity},\ }\href {https://doi.org/10.1364/OL.40.002112} {\bibfield  {journal}
  {\bibinfo  {journal} {Opt. Lett.}\ }\textbf {\bibinfo {volume} {40}},\
  \bibinfo {pages} {2112} (\bibinfo {year} {2015})}\BibitemShut {NoStop}%
\bibitem [{\citenamefont {Matei}\ \emph {et~al.}(2017)\citenamefont {Matei},
  \citenamefont {Legero}, \citenamefont {H\"afner}, \citenamefont {Grebing},
  \citenamefont {Weyrich}, \citenamefont {Zhang}, \citenamefont {Sonderhouse},
  \citenamefont {Robinson}, \citenamefont {Ye}, \citenamefont {Riehle},\ and\
  \citenamefont {Sterr}}]{mat17a}%
  \BibitemOpen
  \bibfield  {author} {\bibinfo {author} {\bibfnamefont {D.~G.}\ \bibnamefont
  {Matei}}, \bibinfo {author} {\bibfnamefont {T.}~\bibnamefont {Legero}},
  \bibinfo {author} {\bibfnamefont {S.}~\bibnamefont {H\"afner}}, \bibinfo
  {author} {\bibfnamefont {C.}~\bibnamefont {Grebing}}, \bibinfo {author}
  {\bibfnamefont {R.}~\bibnamefont {Weyrich}}, \bibinfo {author} {\bibfnamefont
  {W.}~\bibnamefont {Zhang}}, \bibinfo {author} {\bibfnamefont
  {L.}~\bibnamefont {Sonderhouse}}, \bibinfo {author} {\bibfnamefont {J.~M.}\
  \bibnamefont {Robinson}}, \bibinfo {author} {\bibfnamefont {J.}~\bibnamefont
  {Ye}}, \bibinfo {author} {\bibfnamefont {F.}~\bibnamefont {Riehle}},\ and\
  \bibinfo {author} {\bibfnamefont {U.}~\bibnamefont {Sterr}},\ }\bibfield
  {title} {\bibinfo {title} {$1.5~\mu$m lasers with sub-{10 mHz} linewidth},\
  }\href {https://doi.org/10.1103/PhysRevLett.118.263202} {\bibfield  {journal}
  {\bibinfo  {journal} {Phys. Rev. Lett.}\ }\textbf {\bibinfo {volume} {118}},\
  \bibinfo {pages} {263202} (\bibinfo {year} {2017})}\BibitemShut {NoStop}%
\bibitem [{\citenamefont {Riedel}\ \emph {et~al.}(2020)\citenamefont {Riedel},
  \citenamefont {Al-Masoudi}, \citenamefont {Benkler}, \citenamefont
  {D\"orscher}, \citenamefont {Gerginov}, \citenamefont {Grebing},
  \citenamefont {H\"afner}, \citenamefont {Huntemann}, \citenamefont
  {Lipphardt}, \citenamefont {Lisdat}, \citenamefont {Peik}, \citenamefont
  {Piester}, \citenamefont {Sanner}, \citenamefont {Tamm}, \citenamefont
  {Weyers}, \citenamefont {Denker}, \citenamefont {Timmen}, \citenamefont
  {Voigt}, \citenamefont {Calonico}, \citenamefont {Cerretto}, \citenamefont
  {Costanzo}, \citenamefont {Levi}, \citenamefont {Sesia}, \citenamefont
  {Achkar}, \citenamefont {Gu\'ena}, \citenamefont {Abgrall}, \citenamefont
  {Rovera}, \citenamefont {Chupin}, \citenamefont {Shi}, \citenamefont
  {Bilicki}, \citenamefont {Bookjans}, \citenamefont {Lodewyck}, \citenamefont
  {Le~Targat}, \citenamefont {Delva}, \citenamefont {Bize}, \citenamefont
  {Baynes}, \citenamefont {Baynham}, \citenamefont {Bowden}, \citenamefont
  {Gill}, \citenamefont {Godun}, \citenamefont {Hill}, \citenamefont {Hobson},
  \citenamefont {Jones}, \citenamefont {King}, \citenamefont {Nisbet-Jones},
  \citenamefont {Rolland}, \citenamefont {Shemar}, \citenamefont {Whibberley},\
  and\ \citenamefont {Margolis}}]{rie20}%
  \BibitemOpen
  \bibfield  {author} {\bibinfo {author} {\bibfnamefont {F.}~\bibnamefont
  {Riedel}}, \bibinfo {author} {\bibfnamefont {A.}~\bibnamefont {Al-Masoudi}},
  \bibinfo {author} {\bibfnamefont {E.}~\bibnamefont {Benkler}}, \bibinfo
  {author} {\bibfnamefont {S.}~\bibnamefont {D\"orscher}}, \bibinfo {author}
  {\bibfnamefont {V.}~\bibnamefont {Gerginov}}, \bibinfo {author}
  {\bibfnamefont {C.}~\bibnamefont {Grebing}}, \bibinfo {author} {\bibfnamefont
  {S.}~\bibnamefont {H\"afner}}, \bibinfo {author} {\bibfnamefont
  {N.}~\bibnamefont {Huntemann}}, \bibinfo {author} {\bibfnamefont
  {B.}~\bibnamefont {Lipphardt}}, \bibinfo {author} {\bibfnamefont
  {C.}~\bibnamefont {Lisdat}}, \bibinfo {author} {\bibfnamefont
  {E.}~\bibnamefont {Peik}}, \bibinfo {author} {\bibfnamefont {D.}~\bibnamefont
  {Piester}}, \bibinfo {author} {\bibfnamefont {C.}~\bibnamefont {Sanner}},
  \bibinfo {author} {\bibfnamefont {C.}~\bibnamefont {Tamm}}, \bibinfo {author}
  {\bibfnamefont {S.}~\bibnamefont {Weyers}}, \bibinfo {author} {\bibfnamefont
  {H.}~\bibnamefont {Denker}}, \bibinfo {author} {\bibfnamefont
  {L.}~\bibnamefont {Timmen}}, \bibinfo {author} {\bibfnamefont
  {C.}~\bibnamefont {Voigt}}, \bibinfo {author} {\bibfnamefont
  {D.}~\bibnamefont {Calonico}}, \bibinfo {author} {\bibfnamefont
  {G.}~\bibnamefont {Cerretto}}, \bibinfo {author} {\bibfnamefont {G.~A.}\
  \bibnamefont {Costanzo}}, \bibinfo {author} {\bibfnamefont {F.}~\bibnamefont
  {Levi}}, \bibinfo {author} {\bibfnamefont {I.}~\bibnamefont {Sesia}},
  \bibinfo {author} {\bibfnamefont {J.}~\bibnamefont {Achkar}}, \bibinfo
  {author} {\bibfnamefont {J.}~\bibnamefont {Gu\'ena}}, \bibinfo {author}
  {\bibfnamefont {M.}~\bibnamefont {Abgrall}}, \bibinfo {author} {\bibfnamefont
  {G.~D.}\ \bibnamefont {Rovera}}, \bibinfo {author} {\bibfnamefont
  {B.}~\bibnamefont {Chupin}}, \bibinfo {author} {\bibfnamefont
  {C.}~\bibnamefont {Shi}}, \bibinfo {author} {\bibfnamefont {S.}~\bibnamefont
  {Bilicki}}, \bibinfo {author} {\bibfnamefont {E.}~\bibnamefont {Bookjans}},
  \bibinfo {author} {\bibfnamefont {J.}~\bibnamefont {Lodewyck}}, \bibinfo
  {author} {\bibfnamefont {R.}~\bibnamefont {Le~Targat}}, \bibinfo {author}
  {\bibfnamefont {P.}~\bibnamefont {Delva}}, \bibinfo {author} {\bibfnamefont
  {S.}~\bibnamefont {Bize}}, \bibinfo {author} {\bibfnamefont {F.~N.}\
  \bibnamefont {Baynes}}, \bibinfo {author} {\bibfnamefont {C.}~\bibnamefont
  {Baynham}}, \bibinfo {author} {\bibfnamefont {W.}~\bibnamefont {Bowden}},
  \bibinfo {author} {\bibfnamefont {P.}~\bibnamefont {Gill}}, \bibinfo {author}
  {\bibfnamefont {R.~M.}\ \bibnamefont {Godun}}, \bibinfo {author}
  {\bibfnamefont {I.~R.}\ \bibnamefont {Hill}}, \bibinfo {author}
  {\bibfnamefont {R.}~\bibnamefont {Hobson}}, \bibinfo {author} {\bibfnamefont
  {J.~M.}\ \bibnamefont {Jones}}, \bibinfo {author} {\bibfnamefont {S.~A.}\
  \bibnamefont {King}}, \bibinfo {author} {\bibfnamefont {P.}~\bibnamefont
  {Nisbet-Jones}}, \bibinfo {author} {\bibfnamefont {A.}~\bibnamefont
  {Rolland}}, \bibinfo {author} {\bibfnamefont {S.~L.}\ \bibnamefont {Shemar}},
  \bibinfo {author} {\bibfnamefont {P.~B.}\ \bibnamefont {Whibberley}},\ and\
  \bibinfo {author} {\bibfnamefont {H.~S.}\ \bibnamefont {Margolis}},\
  }\bibfield  {title} {\bibinfo {title} {Direct comparisons of {E}uropean
  primary and secondary frequency standards via satellite techniques},\ }\href
  {https://doi.org/10.1088/1681-7575/ab6745} {\bibfield  {journal} {\bibinfo
  {journal} {Metrologia}\ }\textbf {\bibinfo {volume} {57}},\ \bibinfo {pages}
  {045005} (\bibinfo {year} {2020})}\BibitemShut {NoStop}%
\bibitem [{\citenamefont {Denker}\ \emph {et~al.}(2019)\citenamefont {Denker},
  \citenamefont {Timmen},\ and\ \citenamefont {V\"olksen}}]{den19}%
  \BibitemOpen
  \bibfield  {author} {\bibinfo {author} {\bibfnamefont {H.}~\bibnamefont
  {Denker}}, \bibinfo {author} {\bibfnamefont {L.}~\bibnamefont {Timmen}},\
  and\ \bibinfo {author} {\bibfnamefont {C.}~\bibnamefont {V\"olksen}},\ }\href
  {https://doi.org/10.15488/9163} {\emph {\bibinfo {title} {Report on levelling
  and {GNSS} results for stations on the {MPQ} campus in {Garching}}}},\
  \bibinfo {type} {Tech. Rep.}\ (\bibinfo {year} {2019})\BibitemShut {NoStop}%
\bibitem [{\citenamefont {Dick}(1988)}]{dic87}%
  \BibitemOpen
  \bibfield  {author} {\bibinfo {author} {\bibfnamefont {G.~J.}\ \bibnamefont
  {Dick}},\ }\bibfield  {title} {\bibinfo {title} {Local oscillator induced
  instabilities in trapped ion frequency standards},\ }in\ \href
  {https://apps.dtic.mil/sti/citations/ADA502386} {\emph {\bibinfo {booktitle}
  {Proceedings of $19^{th}$ Annu. Precise Time and Time Interval Meeting,
  Redendo Beach, 1987}}},\ \bibinfo {series and number} {\bibinfo {number}
  {ADA502386}}\ (\bibinfo  {publisher} {U.S. Naval Observatory},\ \bibinfo
  {address} {Washington, DC},\ \bibinfo {year} {1988})\ pp.\ \bibinfo {pages}
  {133--147}\BibitemShut {NoStop}%
\bibitem [{\citenamefont {Herbers}\ \emph {et~al.}(2022)\citenamefont
  {Herbers}, \citenamefont {H\"{a}fner}, \citenamefont {D\"{o}rscher},
  \citenamefont {L\"{u}cke}, \citenamefont {Sterr},\ and\ \citenamefont
  {Lisdat}}]{her22}%
  \BibitemOpen
  \bibfield  {author} {\bibinfo {author} {\bibfnamefont {S.}~\bibnamefont
  {Herbers}}, \bibinfo {author} {\bibfnamefont {S.}~\bibnamefont {H\"{a}fner}},
  \bibinfo {author} {\bibfnamefont {S.}~\bibnamefont {D\"{o}rscher}}, \bibinfo
  {author} {\bibfnamefont {T.}~\bibnamefont {L\"{u}cke}}, \bibinfo {author}
  {\bibfnamefont {U.}~\bibnamefont {Sterr}},\ and\ \bibinfo {author}
  {\bibfnamefont {C.}~\bibnamefont {Lisdat}},\ }\bibfield  {title} {\bibinfo
  {title} {Transportable clock laser system with an instability of $1.6 \times
  10^{-16}$},\ }\href {https://doi.org/10.1364/OL.470984} {\bibfield  {journal}
  {\bibinfo  {journal} {Opt. Lett.}\ }\textbf {\bibinfo {volume} {47}},\
  \bibinfo {pages} {5441} (\bibinfo {year} {2022})}\BibitemShut {NoStop}%
\bibitem [{\citenamefont {Katori}(2021)}]{kat21}%
  \BibitemOpen
  \bibfield  {author} {\bibinfo {author} {\bibfnamefont {H.}~\bibnamefont
  {Katori}},\ }\bibfield  {title} {\bibinfo {title} {Longitudinal {Ramsey}
  spectroscopy of atoms for continuous operation of optical clocks},\ }\href
  {https://doi.org/10.35848/1882-0786/ac0e16} {\bibfield  {journal} {\bibinfo
  {journal} {Appl. Phys. Express}\ }\textbf {\bibinfo {volume} {14}},\ \bibinfo
  {pages} {072006} (\bibinfo {year} {2021})}\BibitemShut {NoStop}%
\bibitem [{\citenamefont {Voigt}\ \emph {et~al.}(2016)\citenamefont {Voigt},
  \citenamefont {Denker},\ and\ \citenamefont {Timmen}}]{voi16}%
  \BibitemOpen
  \bibfield  {author} {\bibinfo {author} {\bibfnamefont {C.}~\bibnamefont
  {Voigt}}, \bibinfo {author} {\bibfnamefont {H.}~\bibnamefont {Denker}},\ and\
  \bibinfo {author} {\bibfnamefont {L.}~\bibnamefont {Timmen}},\ }\bibfield
  {title} {\bibinfo {title} {Time-variable gravity potential components for
  optical clock comparisons and the definition of international time scales},\
  }\href {https://doi.org/10.1088/0026-1394/53/6/1365} {\bibfield  {journal}
  {\bibinfo  {journal} {Metrologia}\ }\textbf {\bibinfo {volume} {53}},\
  \bibinfo {pages} {1365} (\bibinfo {year} {2016})}\BibitemShut {NoStop}%
\bibitem [{\citenamefont {Tanaka}\ and\ \citenamefont {Katori}(2021)}]{tan21b}%
  \BibitemOpen
  \bibfield  {author} {\bibinfo {author} {\bibfnamefont {Y.}~\bibnamefont
  {Tanaka}}\ and\ \bibinfo {author} {\bibfnamefont {H.}~\bibnamefont
  {Katori}},\ }\bibfield  {title} {\bibinfo {title} {Exploring potential
  applications of optical lattice clocks in a plate subduction zone},\ }\href
  {https://doi.org/10.1007/s00190-021-01548-y} {\bibfield  {journal} {\bibinfo
  {journal} {J. Geod.}\ }\textbf {\bibinfo {volume} {95}},\ \bibinfo {pages}
  {1} (\bibinfo {year} {2021})}\BibitemShut {NoStop}%
\bibitem [{\citenamefont {Takamoto}\ \emph {et~al.}(2022)\citenamefont
  {Takamoto}, \citenamefont {Tanaka},\ and\ \citenamefont {Katori}}]{tak22}%
  \BibitemOpen
  \bibfield  {author} {\bibinfo {author} {\bibfnamefont {M.}~\bibnamefont
  {Takamoto}}, \bibinfo {author} {\bibfnamefont {Y.}~\bibnamefont {Tanaka}},\
  and\ \bibinfo {author} {\bibfnamefont {H.}~\bibnamefont {Katori}},\
  }\bibfield  {title} {\bibinfo {title} {A perspective on the future of
  transportable optical lattice clocks},\ }\href
  {https://doi.org/10.1063/5.0087894} {\bibfield  {journal} {\bibinfo
  {journal} {Appl. Phys. Lett.}\ }\textbf {\bibinfo {volume} {120}},\ \bibinfo
  {pages} {140502} (\bibinfo {year} {2022})}\BibitemShut {NoStop}%
\bibitem [{\citenamefont {Benkler}\ \emph {et~al.}(2020)\citenamefont
  {Benkler}, \citenamefont {Lipphardt}, \citenamefont {Puppe}, \citenamefont
  {Wilk}, \citenamefont {Rohde},\ and\ \citenamefont {Sterr}}]{ben20}%
  \BibitemOpen
  \bibfield  {author} {\bibinfo {author} {\bibfnamefont {E.}~\bibnamefont
  {Benkler}}, \bibinfo {author} {\bibfnamefont {B.}~\bibnamefont {Lipphardt}},
  \bibinfo {author} {\bibfnamefont {T.}~\bibnamefont {Puppe}}, \bibinfo
  {author} {\bibfnamefont {R.}~\bibnamefont {Wilk}}, \bibinfo {author}
  {\bibfnamefont {F.}~\bibnamefont {Rohde}},\ and\ \bibinfo {author}
  {\bibfnamefont {U.}~\bibnamefont {Sterr}},\ }\bibfield  {title} {\bibinfo
  {title} {End-to-end topology for fiber comb based optical frequency transfer
  at the $10^{-21}$ level: erratum},\ }\href
  {https://doi.org/10.1364/OE.27.036886} {\bibfield  {journal} {\bibinfo
  {journal} {Opt. Express}\ }\textbf {\bibinfo {volume} {28}},\ \bibinfo
  {pages} {15023} (\bibinfo {year} {2020})}\BibitemShut {NoStop}%
\end{thebibliography}

%

\end{document}


\newcommand{\e}[1]{\times 10^{#1}}


\title{Supplemental Material for Long-distance chronometric leveling with a portable optical clock}

\author{J.~Grotti}
\author{I.~Nosske}
\author{S.~B.~Koller}
\author{S.~Herbers}
\affiliation{Physikalisch-Technische Bundesanstalt, Bundesallee 100, 38116 
	Braunschweig, Germany}

\author{H.~Denker}
\author{L.~Timmen}
\affiliation{Institut f\"ur Erdmessung, Leibniz Universit\"at Hannover (LUH), Schneiderberg 50, 30167 Hannover, Germany}

\author{G.~Vishnyakova}
\affiliation{Max-Planck-Institut f\"ur Quantenoptik, Hans-Kopfermann-Stra{\ss}e 1, 85748 Garching, Germany}

\author{G.~Grosche}
\author{T.~Waterholter}
\author{A.~Kuhl}
\author{S.~Koke}
\author{E.~Benkler}
\affiliation{Physikalisch-Technische Bundesanstalt, Bundesallee 100, 38116 Braunschweig, Germany}

\author{M.~Giunta}
\affiliation{Max-Planck-Institut f\"ur Quantenoptik, Hans-Kopfermann-Stra{\ss}e 1, 85748 Garching, Germany}
\affiliation{Menlo Systems GmbH, Bunsenstra{\ss}e 5, 82152 Martinsried, Germany}

\author{L.~Maisenbacher}
\author{A.~Matveev}
\affiliation{Max-Planck-Institut f\"ur Quantenoptik, Hans-Kopfermann-Stra{\ss}e 1, 85748 Garching, Germany}

\author{S.~D\"orscher}
\author{R.~Schwarz}
\author{A.~Al-Masoudi}
\affiliation{Physikalisch-Technische Bundesanstalt, Bundesallee 100, 38116 Braunschweig, Germany}

\author{T.~W.~H\"ansch}
\author{Th.~Udem}
\affiliation{Max-Planck-Institut f\"ur Quantenoptik, Hans-Kopfermann-Stra{\ss}e 1, 85748 Garching, Germany}

\author{R.~Holzwarth}
\affiliation{Max-Planck-Institut f\"ur Quantenoptik, Hans-Kopfermann-Stra{\ss}e 1, 85748 Garching, Germany}
\affiliation{Menlo Systems GmbH, Bunsenstra{\ss}e 5, 82152 Martinsried, Germany}

\author{C.~Lisdat}
\email{christian.lisdat@ptb.de}
\affiliation{Physikalisch-Technische Bundesanstalt, Bundesallee 100, 38116 Braunschweig, Germany}

\date{\today}
\maketitle

\section{Consideration of correlations in weighted averaging}\label{sec:cor}
According to Eq.~(16) of Ref. \cite{gum08}, the standard uncertainty of a measured value $y$, being the result of a function $f(x_1,...,x_N)$ dependent on several input estimates $x_i$ with $i = 1...N$, in presence of correlations between the $x_i$ is expressed by
\begin{eqnarray}
	&& u(y) = \left[ \sum_{i=1}^N \left( \frac{\partial f}{\partial x_i} \right)^2 u^2(x_i)  \right.  \\
	&&+ \left. 2 \sum_{i=1}^{N-1} \sum_{j=i+1}^{N} \, \left( \frac{\partial f}{\partial x_i} \right) \left( \frac{\partial f}{\partial x_j} \right)   u(x_i) u(x_j) \, r(x_i,x_j) \right]^{1/2}. \nonumber
	\label{eq:gum}
\end{eqnarray}

The correlations are described for pairs of input estimates by coefficients $r(x_i,x_j)$ ranging from +1 (fully correlated) through 0 (uncorrelated) to $-1$ (fully anti-correlated). 
$u(x_i)$ denotes the uncertainty of the input $x_i$.

In the present case, $f$ is the weighted average of the mean daily frequency offsets between the clocks as described by Eqs.~(1) and (2) in the main paper:
\begin{eqnarray}
	f = \overline{\Delta \nu} = \frac{\sum_\mathrm{day} \Delta \nu_\mathrm{day} / u^2(\Delta \nu_\mathrm{day})   }{\sum_\mathrm{day'} 1/ u^2(\Delta \nu_\mathrm{day'})  }
	\label{eq:mean}
\end{eqnarray}
\noindent where we have omitted the labels rem/loc for clarity.
With the definitions for $\Delta \nu_\mathrm{day}$ from the main paper and the described correlations between the respective contributions, Eq.~(\ref{eq:mean}) factorizes such that $r(x_i,x_j)$ are either 0 or +1. 
The partial derivatives are given by
\begin{eqnarray}
	\frac{\partial f}{\partial \Delta \nu_\mathrm{day}} = \frac{1/u^2(\Delta \nu_\mathrm{day})}{\sum_\mathrm{day'} 1/ u^2(\Delta \nu_\mathrm{day'})}.
	\label{eq:deriv}
\end{eqnarray}
Similarly, instead of calculating the uncertainty of the average of a local or remote campaign, also the uncertainty of the chronometric potential difference $\Delta U_\mathrm{chron}$ in Eq.~(3) can be evaluated.

\section{Uncertainties of the transportable lattice clock Sr2}\label{sec:unc}

Here, we give an overview on the evaluation of uncertainties of the transportable lattice clock Sr2.

\subsection {Ambient BBR shift}
The procedure remains unchanged from the one described in Ref.~\cite{kol17}. 
We measure the temperature of the vacuum system with eight Pt100 sensors on and in the vacuum chamber. 
The sensors are placed at the regions with highest and lowest temperature to provide a realistic estimate of the temperature span over the chamber. 
We estimate the representative temperature by $(T_\mathrm{max} - T_\mathrm{min})/2$ and its uncertainty by $(T_\mathrm{max} - T_\mathrm{min})/\sqrt{12}$.
Here, it is assumed that the representative temperature that leads to the BBR shift experienced by the atoms lies with equal probability between the boundaries $T_\mathrm{max}$ and $T_\mathrm{min}$ \cite{gum08}. The atomic coefficients for the BBR correction are taken from \cite{mid12a} and \cite{nic15}.
We did not update the data analysis with the corrected dynamic polarizability \cite{lis21a} because the change in the BBR correction is to a very large extent common mode in all measurements and hence does not influence the final result.

\subsection {Oven BBR}
During interrogation, the atoms in the lattice have been shielded from BBR from the hot atomic oven by a low-emissivity shutter. 
This procedure leads to a strong reduction of the possible effect and thus a negligible uncertainty contribution (see also Sr1 in Tab.~1 for the uncertainty without beam shutter). 

\subsection {Lattice light shift}
Following the lattice light shift model in Ref.~\cite{ush18}, we describe first- and higher-order lattice light shifts in a combined model.
For the determination of the residual lattice light shift, we interleave during normal clock operation measurements \cite{alm15} with lattice depths of typically 75~$E_r$ and 170~$E_r$ and measure the differential frequency shift.
$E_r$ is the lattice photon recoil energy.

\subsection {Probe light shift and line pulling}
The $^{87}$Sr clock transition was typically interrogated with Rabi pulses of 180~ms length.
With a lifetime of the excited clock state of about 120~s \cite{boy07a,doe18, mun21} and the light shift coefficient of about $-13~\mathrm{Hz \, cm^2 \, W^{-1}}$ \cite{bai07}, we estimate a negligible clock laser induced light shift of few $10^{-19}$.

During clock operation, we interrogate the $\Delta m_F = 0$ $|m_F|=9/2$ transitions in a homogeneous magnetic bias field that causes a splitting of the two transitions from $m_F = \pm 9/2$ of about 420~Hz. 
To reduce line pulling effects, we remove population from other Zeeman levels by resonant laser radiation on the 461~nm singlet transition after we have excited the population in $m_F=\pm 9/2$ from the ground state to the upper clock state by a short $\pi$-pulse in a large bias field. 
The actual clock interrogation is then performed by de-excitation to the ground state.
Further line pulling can occur by off-resonantly driven $\sigma$ transitions. 
We checked for remaining population in $m_F \neq \pm 9/2$ and weak $\sigma$ transitions and estimate that their remaining influences will cause shifts below the $10^{-19}$-level.

\subsection {Second order Zeeman shift}
With the lock scheme to the transitions from the $m_F = \pm 9/2$ states, we obtain online information on the magnetic field strength at the position of the atoms provided that the lattice vector shift \cite{shi15} is sufficiently small. 
During lattice light shift determinations, the vector light shift was found to be negligible compared to the typical daily variation of $\pm 3$~Hz of the line splitting caused by changes in the ambient and bias field. 
The latter is the main source of uncertainty for operation at a splitting of about 420~Hz between the interrogated transitions. 
The atomic correction coefficient is well-known \cite{bot19} and does not significantly contribute to the estimated uncertainty of $2 \times 10^{-18}$. 
This contribution can be reduced further by a time-resolved correction of the second-order Zeeman shift.

\subsection {Cold and background gas collisions}
We assess the influence of cold collisions by a variation of the number of interrogated atoms in an interleaved measurement and observation of the related frequency shift. 
We observe zero-compatible shifts and estimate the residual uncertainty based on the strength of the detected signal.

The shift by collisions with the background gas is derived from the observed trap lifetime of $1.55(26) \, \mathrm{s}$ and the shift coefficient from Ref.~\cite{alv19}. Here, we assume that our background gas composition is dominated by hydrogen. We find a shift of $1.9(4) \times 10^{-17}$.

\subsection {Servo error}
During the clock comparisons at PTB, the interrogation laser of Sr2 was phase-locked to the Sr1 clock laser system to improve its short-term stability and reduce averaging times.
Since the Sr1 laser is tied to the strontium transition frequency during the valid times of the comparison, drifts are essentially removed from the Sr2 interrogation laser.
Consequently, we set the lock error to zero for these measurements.

For the remote campaign at MPQ, we estimate the possible lock error from the gains of the servo loops for frequency and drift correction and from the maximum drift rate change observed during the measurements \cite{fal11}. 
We use the maximum servo error of typically $2.2 \times 10^{-18}$ as uncertainty.

\subsection {Tunneling}
The lattice of Sr2 is oriented under an angle of $50^\circ$ with respect to gravity, which results in a potential difference between neighbouring lattice sites.
In combination with a deep lattice of 50~$E_r$ and more, tunneling is strongly suppressed as are tunneling-related frequency shifts \cite{lem05}. 

\subsection {DC Stark shift}
The DC Stark shift is estimated from the distance of 54~mm between two windows that are the surfaces closest to the atoms \cite{kol17}.
On the inner side, these windows are coated with a conducting material (indium tin oxide).
Under these conditions, patch potentials should be below 100~mV, inducing a DC Stark shift of less than $-1\times 10^{-19}$ \cite{mid12a}.

 \begin{figure}[b]
    \includegraphics{blank.png}
\end{figure}


%

%
\bibliographystyle{apsrev4-2}